\documentclass[%
 aip,
 amsmath,amssymb,
 reprint,%
]{revtex4-1}

\usepackage{graphicx}
\usepackage{dcolumn}
\usepackage{bm}

\usepackage[utf8]{inputenc}
\usepackage[T1]{fontenc}
\usepackage{mathptmx}
\usepackage{etoolbox}

\usepackage{xcolor}

\makeatletter
\def\@email#1#2{%
 \endgroup
 \patchcmd{\titleblock@produce}
  {\frontmatter@RRAPformat}
  {\frontmatter@RRAPformat{\produce@RRAP{*#1\href{mailto:#2}{#2}}}\frontmatter@RRAPformat}
  {}{}
}%
\makeatother

\usepackage{mathtools}
\DeclarePairedDelimiter\bra{\langle}{\rvert}
\DeclarePairedDelimiter\ket{\lvert}{\rangle}
\DeclarePairedDelimiterX\braket[2]{\langle}{\rangle}{#1 \delimsize\vert #2}
\newcommand{\absDeriv}[1]{\frac{\mathrm{d}}{\mathrm{d}#1}}

\begin{document}

\preprint{AIP/123-QED}

\title[Modelling HHG in QDs using a real-space tight-binding approach]{Modeling high-order harmonic generation in quantum dots using a real-space tight-binding approach}
\author{Martin Th\"ummler}
\affiliation{%
    Institute of Physical Chemistry, Friedrich Schiller University Jena, Germany
}%
\author{Alexander Croy}%
\affiliation{%
    Institute of Physical Chemistry, Friedrich Schiller University Jena, Germany
}%

\author{Ulf Peschel}
\affiliation{
    Institute of Condensed Matter Theory and Optics, Friedrich Schiller University Jena, Germany 
}%
\affiliation{
    Abbe School of Photonics, Friedrich Schiller University Jena, Germany
}
\author{Stefanie Gr\"afe}
\affiliation{%
    Institute of Physical Chemistry, Friedrich Schiller University Jena, Germany
}
\affiliation{
    Abbe School of Photonics, Friedrich Schiller University Jena, Germany
}
\affiliation{%
    Fraunhofer Institute for Applied Optics and Precision Engineering, Jena, Germany
}%

\date{\today}

\begin{abstract}
Recently, the size-dependence of high-order harmonic generation (HHG) in quantum dots has been investigated experimentally.
In particular, for longer driving wavelengths and QDs smaller than 3\,nm, HHG was strongly suppressed, however, there is no computational model capable of describing the strong-field response of such systems.
In this work, we introduce a computationally efficient three-dimensional real-space tight-binding model specifically designed for the simulation of HHG in confined systems.
The model parameters are meticulously derived from density functional theory (DFT) calculations for the semiconductor bulk, followed by a process of Wannierization.
Our findings demonstrate that the proposed model accurately captures the observed dependency of the HHG yield on the quantum dot size.
Additionally, we simulate the HHG yield for elliptically polarized pulses for different QD-sizes and driving wavelengths up to $5\,\mu{\mathrm{m}}$.
The herein proposed model fills the theoretical void in simulating HHG within medium-sized nanostructures, which cannot be described by methods applied for periodic solids or small molecules or atoms.
\end{abstract}

\maketitle

\section{\label{sec:intro}Introduction}

The interaction of intense laser fields with matter gives rise to many nonlinear optical phenomena, among them the generation of high-order harmonics (HHG).
HHG has been extensively studied in various systems, initially in atomic and molecular gases \cite{ferray_1988}, and more recently, in solid-state materials \cite{ghimire_2011,li_resonance_2023} and even in liquid jets \cite{luu_extremeultraviolet_2018}.
Very recently, HHG has been experimentally investigated in a different class of materials: semiconductor quantum dots (QD) \cite{nakagawa_2022,gopalakrishna_2023}, featuring discrete energy levels and size-dependent electronic properties.
The quantum confinement in QDs leads to unique electronic behaviors that differ significantly from their bulk counterparts. 
Experimentally, an unexpected strong suppression of HHG in dots sizes of less than $3\,\mathrm{nm}$ has been observed.
However, a comprehensive theoretical framework that adequately describes HHG in these nanostructures, particularly one that accounts for their specific characteristics and size dependencies, has been lacking: with QD consisting of 100s of atoms, commonly used numerical methods to describe HHGs in molecules or atoms, such as the direct solution of the Schrödinger equation \cite{reiff2020single} or real-space real-time time-dependent density functional theory (rt-TDDFT) calculations \cite{chu_2001,tancogne2018atomic}, cannot be applied because of their high computational effort.
On the other side, HHG in periodic structures is commonly described by the semiconductor-Bloch-equations (SBEs) \cite{lindberg_1988,silva_2019}, tight-binding models \cite{jiang_2018}, or TDDFT calculations \cite{tancogne-dejean_2017_1}.
But these periodic models cannot account for the finite size of the QDs.

In this work, we present a real-space tight-binding model in three-dimensions tailored for the simulation of HHG in QDs.
It extends our previous one-dimensional model with cosine-shaped bands \cite{peschel_2022} and incorporates parameters derived from density functional theory (DFT) calculations for the corresponding bulk, followed by subsequent representation by maximally localized Wannier functions \cite{marzari_1997,marzari_2012}.
These allow to account for the finite spatial structure of the QDs while naturally including the periodic limit.
Our finite-size model is computationally feasible for small to medium sized (up to 500 atoms) QDs.
It predicts the HHG yield for very small structures in good agreement with rt-TDDFT calculations. To contrast our approach to the periodic limit, we also developed a real-space periodic model, which we compare with rt-TDDFT calculations.
Our real-space periodic model is strongly related to the SBEs in the Wannier moving-frame basis \cite{silva_2019,thummler_semiconductor_2025} and thus allows for a direct comparison with this well-established approach \cite{yue_2022}.
In this study, we consider CdSe QDs for our numerical investigations, see fig. 1 for some of the geometries considered.
We apply our three-dimensional finite-size model to recover the experimentally observed QD size-dependence of the HHG yield \cite{nakagawa_2022,gopalakrishna_2023}.
Moreover, we simulate the dependence of the HHG yield on the ellipticity of the driving pulse, which we find to show a behavior similar to solids \cite{liu_2016,nano11010004}.

This model not only facilitates a deeper understanding of size-related effects on HHG in quantum dots but also enables the exploration of HHG responses to varying pulse shapes and wavelengths. The significance of our work lies not only in its contribution to the theoretical understanding of HHG in nano-structured materials but also in its potential to guide experimental studies, allowing to analyze size and shape dependencies of HHG yields. 

\begin{figure}
    \includegraphics[width=0.85\linewidth, trim=2cm 0cm 1cm 0cm]{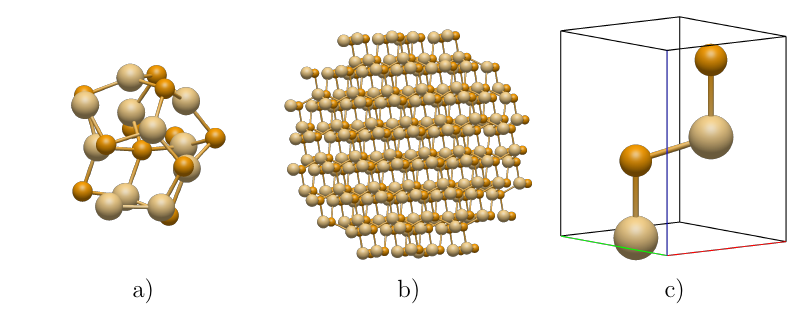}
    \caption{Examples of considered geometries: a) Optimized $\mathrm{Cd}_{11}\mathrm{Se}_{11}$ cluster (used in rt-TDDFT calculations) b) $\mathrm{Cd}_{198}\mathrm{Se}_{198}$ quantum dot with diameter $2.8\,\mathrm{nm}$ (used in our real-space model) , and
    c) unit cell of CdSe (wurtzite).
    }
    \label{fig:geometries}
\end{figure}

This work is organized as follows: In section \ref{sec:theory}, we first discuss the theoretical background of our model and subsequently our finite-size model in section \ref{sec:finiteModel}, where we also describe the implementation details for solving the equations of motion numerically.
We introduce our the real-space periodic model in section \ref{sec:periodicModel}.
In section \ref{sec:numeric}, we first compare our models to rt-TDDFT calculations, where computationally feasible.
Additionally, we approve our real-space periodic model based on absorption spectra calculations.
Finally, we present our simulation results for the QD-size and the ellipticity dependence and conclude in section \ref{sec:conclusion}.

\section{\label{sec:theory}Theoretical background}
Throughout this work, atomic units are employed, unless stated otherwise.
We start with the time-dependent Hamiltonian in length gauge
\begin{equation} \label{eq:HinLG}
    \hat{H}(t) = \hat{H}_0 + \mathbf{E}(t) \cdot\hat{\mathbf{r}}
\end{equation}
with the field-free Hamiltonian $\hat{H}_0$ and the time-dependent electric field $\mathbf{E}(t)$.
The latter is defined via the vector potential $\mathbf{A}$ as 
\begin{align}
    \mathbf{E} = -\frac{\partial}{\partial t} \mathbf{A}
.
\end{align}

We represent Hamiltonian $\hat{H}_0$ of the QD in the Wannier basis \cite{silva_2019,marzari_2012}.
Employing the Wannier functions satisfies the periodic limit naturally, as they are obtained from Bloch functions.
Additionally, they are localized \cite{marzari_1997} in real-space allowing to capture finite size effects accurately with a small number of functions per unit cell and lastly, the Hamiltonian and dipole matrix elements are readily available after the Wannierization procedure \cite{marzari_2012}. 
We denote the Bloch-functions (eigenfunctions of the field-free periodic Hamiltonian $\hat{H}_0$) as $\ket{\psi_{n\mathbf{k}}}$ with wave vector $\mathbf{k}$ and band index $n$.
The transformation to Wannier functions \cite{marzari_1997} centered at unit cell $\mathbf{R}$ is given by
\begin{equation}
    \ket{\mathbf{R}n} = \frac{V}{(2\pi)^3} \int \mathrm{d}\mathbf{k} \exp\left(-\mathrm{i}\mathbf{k}\cdot\mathbf{R} \right) \sum\limits_{m} U^\mathbf{k}_{mn}\ket{\psi_{m\mathbf{k}}}
,
\end{equation}
where $n$ is the Wannier function index and $V$ is the unit cell volume.
The semi-unitary matrices $U^{\mathbf{k}}$ are optimized to minimize the spatial extent of the Wannier functions \cite{marzari_1997,pizzi_2020,souza_2001}.
We employ two subsets of the Bloch functions for the Wannierization, one including all valence bands, and the other for all relevant conduction bands.
This way, we obtain orthogonal Wannier functions $\ket{(\mathbf{R}\mu)_\mathrm{c}}$ and $\ket{(\mathbf{R}\mu)_\mathrm{v}}$ for the conduction and valence bands, respectively.
The Hamiltonian $\hat{H}_0$ is expressed in terms of the resulting Wannier functions to represent the conduction bands via
\begin{equation} \label{eq:defWannierH}
    H^{\mathrm{c}}_{\mathbf{R},\mu\nu} = \bra{(\mathbf{0}\mu)_{\mathrm{c}}} \hat{H}_0 \ket{(\mathbf{R}\nu)_\mathrm{c}}
\end{equation}
and analogously $H^{\mathrm{v}}_{\Delta\mathbf{R},\mu\nu}$ for the valence bands.
As our choice of Wannier functions does not mix conduction and valence bands, we have $\bra{(\mathbf{0}\mu)_\mathrm{c}} \hat{H}_0 \ket{(\mathbf{R}\nu)_\mathrm{v}} = 0$.
Due to the periodicity of the lattice, the Hamiltonian matrix elements depend only on the unit cell distance between the Wannier functions, i.e.
\begin{equation}
  \bra{(\mathbf{R}_1\mu)_{\mathrm{c}}} \hat{H}_0 \ket{(\mathbf{R}_2\nu)_\mathrm{c}} = H^{\mathrm{c}}_{\mathbf{R}_2-\mathbf{R}_1,\mu\nu}
\end{equation}
and analogously for the valence bands.
The dipole matrix elements
\begin{equation} \label{eq:defWannierR}
    \mathbf{D}^{\mathrm{cv}}_{\mathbf{R},\mu\nu} = \bra{(\mathbf{0}\mu)_\mathrm{c}} \hat{\mathbf{r}} \ket{(\mathbf{R}\nu)_\mathrm{v}}
\end{equation}
between the Wannier functions can be calculated according to Ref.\ \cite{wang_ab_2006}.
The matrices for $\mathbf{D}^{\mathrm{vv}}$, $\mathbf{D}^{\mathrm{vc}}$, and $\mathbf{D}^{\mathrm{cc}}$ are defined analogously.
The position operator breaks the translational invariance, which manifests as
\begin{align}\label{eq:WannierPosShift}
\bra{(\mathbf{R}_1\mu)_\mathrm{v}} \hat{\mathbf{r}} \ket{(\mathbf{R}_2\nu)_\mathrm{v}} &= 
\delta_{\mathbf{R}_1\mathbf{R}_2}\delta_{\mu\nu}\mathbf{R}_1 + \mathbf{D}^{\mathrm{vv}}_{\mathbf{R}_2-\mathbf{R}_1,\mu\nu}
    ,
\end{align}
and analogously for $\bra{(\mathbf{R}_1\mu)_\mathrm{c}} \hat{\mathbf{r}} \ket{(\mathbf{R}_2\nu)_\mathrm{c}}$, where we introduced the Kronecker $\delta$-symbol.
Contrary, the dipole matrix elements between valence and conduction bands obey
\begin{align}
    \bra{(\mathbf{R}_1\mu)_\mathrm{c}} \hat{\mathbf{r}} \ket{(\mathbf{R}_2\nu)_\mathrm{v}} &= 
\mathbf{D}^{\mathrm{cv}}_{\mathbf{R}_2-\mathbf{R}_1,\mu\nu}
.
\end{align}
In the following, we introduce two different models:
The first one in section \ref{sec:finiteModel} we describes the finite-size QD,
whereas the second one, see section \ref{sec:periodicModel}, includes periodic boundary conditions in real-space to allow for a computational tractable comparison of our model with established methods for periodic systems.

\section{\label{sec:finiteModel}Finite-size model}
The finite-size model and its implementation is the core of our study.
We describe the QD geometry by an assembly of unit cells and setup a tight-binding model via the corresponding Wannier functions, where we do not account for the Coulomb interaction explicitly.
We note, that for nanometer-sized QDs, the exciton binding energies can reach several 100\,meV \cite{meulenberg_determination_2009}.
The energy shifts due to confinement and due to the incident electric field are almost one order of magnitude higher and it is therefore no surprise that experimentally observed HHG spectra of QDs lack pronounced excitonic features.
Moreover, no damage has been observed in those experiments, and thus ionization effects can be assumed to be negligible.
Therefore, we do not describe ionization effects, but note that our model could be extended like in Ref.\ \cite{peschel_2022}, at a high computational cost, because of the large real-space volume required to cover the free-space electron quiver radius \cite{corkum_plasma_1993,tao_photo-electron_2012}.

\subsection{Model Hamiltonian}
We define our real-space tight-binding model in terms of electron ($\hat{e}^\dagger_{\mathbf{R}\mu}$) and hole ($\hat{h}^\dagger_{\mathbf{R}\mu}$) creation operators, which create an electron in the Wannier state $\ket{(\mathbf{R}\mu)_\mathrm{c}}$ from the conduction band and a hole in the state $\ket{(\mathbf{R}\mu)_\mathrm{v}}$ from the valence band, respectively.
Hence, the model Hamiltonian reads
\begin{eqnarray} \label{eq:H0}
    \hat{H}_0 = \sum\limits_{\mathbf{R},\Delta\mathbf{R}} \left[
        \sum\limits_{\mu\nu} H^\mathrm{c}_{\Delta \mathbf{R}, \mu\nu} \hat{e}^\dagger_{\mathbf{R}, \mu} \hat{e}_{\mathbf{R} + \Delta \mathbf{R}, \nu} 
        \right. \nonumber \\ \left.
       + 
        \sum\limits_{\mu\nu} H^\mathrm{v}_{\Delta \mathbf{R}, \mu\nu} \hat{h}^\dagger_{\mathbf{R}, \mu} \hat{h}_{\mathbf{R} + \Delta \mathbf{R}, \nu}
                \right]
.
\end{eqnarray}
The sum contains exactly those terms, where the lattice vectors $\mathbf{R}$ and $\mathbf{R}+\Delta \mathbf{R}$ are both inside a given dot radius.
In other words, the QD geometry is defined as a spherical cut-out of the periodic structure with cuts at unit cell boundaries and not between arbitrary atoms, see fig.~\ref{fig:geometries}b for an example.
As an extension, surface relaxation effects can be included via the Harrison scaling \cite{harrison2012electronic}, as we already considered in our previous one-dimensional model \cite{peschel_2022}.

By using separate sets of Wannier functions for the conduction and valence bands, we avoid a numerical search or diagonalization to find the QD ground state $\ket{\Psi_0}$ which corresponds to completely filled valence bands and empty conduction bands.
The interaction with an optical field, see eq.~\eqref{eq:HinLG}, necessitates also $\hat{\mathbf{r}}$ to be expressed in terms of the electron and hole operators as 
\begin{eqnarray} \label{eq:HD}
    \hat{\mathbf{r}} =& \sum\limits_{\mathbf{R},\Delta\mathbf{R}} \left[
        \sum\limits_{\mu\nu} \mathbf{D}^\mathrm{cc}_{\Delta \mathbf{R}, \mu\nu} \hat{e}^\dagger_{\mathbf{R}, \mu} \hat{e}_{\mathbf{R} + \Delta \mathbf{R}, \nu}
        \right. \nonumber \\
        &\left.
        + \sum\limits_{\mu\nu} \mathbf{D}^\mathrm{vv}_{\Delta \mathbf{R}, \mu\nu} \hat{h}^\dagger_{\mathbf{R},\mu} \hat{h}_{\mathbf{R} + \Delta \mathbf{R}, \nu}
        \right. \nonumber \\
      &\left. 
        \quad\quad\quad\,\,\, + \sum\limits_{\mu\nu} ( 
                        \mathbf{D}^{\mathrm{cv}}_{\Delta \mathbf{R}, \mu\nu} \hat{e}^\dagger_{\mathbf{R}\mu} \hat{h}^\dagger_{\mathbf{R} + \Delta \mathbf{R},\nu} + \text{h.c.})
        \right] \nonumber \\
      &
        + \sum\limits_{\mathbf{R}} \mathbf{R} \left[ \sum\limits_{\mu} \hat{e}^\dagger_{\mathbf{R},\mu} \hat{e}_{\mathbf{R},\mu} - \sum\limits_{\mu} \hat{h}_{\mathbf{R},\mu}^\dagger \hat{h}_{\mathbf{R},\mu} \right]
\end{eqnarray}
with the dipole matrix elements $\mathbf{D}^\mathrm{cc}$ ($\mathbf{D}^\mathrm{vv}$) connected to the electron (hole) states.
Here, $\mathbf{D}^\mathrm{cv}$ describes the transition dipole moment for the creation of an electron-hole pair.
The last line in eq.~\eqref{eq:HD} accounts for the unit cell offset of the centers of the Wannier functions, see eq.~\eqref{eq:WannierPosShift}.

\subsection{Numerical propagation}
We propagate the density matrix operator
\begin{equation}
\hat{\rho} =
\begin{pmatrix}
\hat{n}^\mathrm{e} & \hat{p} \\
\hat{p}^\dagger & \hat{n}^{\mathrm{h}}
\end{pmatrix}
,
\end{equation}
where $\hat{n}^{\mathrm{e}}$, $\hat{p}$, and $\hat{n}^\mathrm{h}$ are block matrices defined as
$\hat{n}^\mathrm{e}_{\mathbf{R}_1\mu\mathbf{R}_2\nu} = \hat{e}^\dagger_{\mathbf{R}_1\mu}\hat{e}_{\mathbf{R}_2\nu}$,
$\hat{p}_{\mathbf{R}_1\mu\mathbf{R}_2\nu} = \hat{h}_{\mathbf{R}_1\mu} \hat{e}_{\mathbf{R}_2\nu}$,
and
$\hat{n}^\mathrm{h}_{\mathbf{R}_1\mu\mathbf{R}_2\nu} = \hat{h}^\dagger_{\mathbf{R}_1\mu}\hat{h}_{\mathbf{R}_2\nu}$ describing the electrons in the conduction bands, the polarization between electrons and holes, and the holes, respectively.
Its time derivative is given by the von-Neumann equation as
\begin{align} \label{eq:finiteSizeDensRaw}
  \mathrm{i}  \absDeriv{t}\hat{\rho} &= \left[\hat{H}, \hat{\rho}\right] = \left[\hat{H}_0, \hat{\rho}\right] + \mathbf{E} \left[\hat{\mathbf{r}}, \hat{\rho}\right]
.
\end{align}
We define the Liouvillian $\hat{L}$ based on the commutator with the Hamiltonian via $\hat{L}\hat{\rho} \equiv [\hat{H}, \hat{\rho}]$.
For our numerical propagation we split the Liouvillian as $\hat{L} = \hat{L}_0 + \mathbf{E} \hat{\mathbf{L}}_\mathrm{D}$ with $\hat{L}_0 \equiv [\hat{H}_0, \hat{\rho}]$, where $\hat{H}_0$ is defined in eq.~\eqref{eq:H0}, and $\hat{\mathbf{L}}_\mathbf{r} \equiv [\hat{\mathbf{r}}, \hat{\rho}]$, with $\hat{\mathbf{r}}$ defined in eq.~\eqref{eq:HD}.
Based on these definitions, we rewrite eq.~\eqref{eq:finiteSizeDensRaw} as
\begin{equation} \label{eq:densProp}
    \mathrm{i}\absDeriv{t}{\hat{\rho}} = \left(\hat{L}_{0} + \mathbf{E}(t) \hat{\mathbf{L}}_{\mathbf{r}}\right)\hat{\rho}
.
\end{equation}
As we neglect the Coulomb interaction, we can evaluate the Liouvillian algebraically in terms of the two-particle correlations $\hat{n}^{\mathrm{e}}$, $\hat{p}$ and $\hat{n}^{\mathrm{h}}$.
The time derivative of the density matrix $\rho = \bra{\Psi_0} \hat{\rho} \ket{\Psi_0}$ is  completely described by the expectation values 
$\bra{\Psi_0}\hat{n}^\mathrm{e}_{\mathbf{R}_1\mu\mathbf{R}_2\nu}\ket{\Psi_0}$,
$\bra{\Psi_0}\hat{p}_{\mathbf{R}_1\mu\mathbf{R}_2\nu} \ket{\Psi_0}$
, and $\bra{\Psi_0}\hat{n}^\mathrm{h}_{\mathbf{R}_1\mu\mathbf{R}_2\nu} \ket{\Psi_0}$. 
For our numerical propagation, we compile these expectation values into a vector and represent the Liouvillian as a sparse matrix acting on that vector.
Since our initial state $\ket{\Psi_0}$ at time $t_0$ contains no excited carriers, the entries of the vector representing the density matrix obey $\rho(t_0)\equiv 0$.
After setting up the required matrix representations $L_0$ and $\mathbf{L}_{\mathbf{r}}$ of the Livovillians, we numerically propagate the vectorized density matrix on a GPU using the Runge-Kutta 4/5 algorithm from the C++ boost odeint library \cite{ahnert_2011} in combination with vexcl \cite{demidov_2013}.

We want to highlight two implementation details, as a naive implementation of the vectorized evaluation of eq.~\eqref{eq:densProp} is more than one order of magnitude slower than our algorithm:
First, we avoid unnecessary entries in the vectorized density matrix by omitting all correlations that are determined (through complex conjugation) by the self-adjointness of the density matrix, that is, $\hat{n}^{\mathrm{e/h}}_{\mathbf{R}_1\mu\mathbf{R}_2\nu} = (\hat{n}^{\mathrm{e/h}}_{\mathbf{R}_2\nu\mathbf{R}_1\mu})^\dagger$.
Second, we do not calculate the sparse matrix representation of $\mathbf{E}\mathbf{L}_{\mathbf{r}}$ explicitly.
Instead, we exploit that the number of Wannier matrix elements is small compared to the size of the Liouvillian matrix, which allows us to efficiently store $\mathbf{E}\mathbf{L}_\mathbf{r}$ as a sparse matrix containing only references to its algebraically different values.
Those values are calculated solely once for each $\mathbf{E}$ field as a function of the Wannier matrix elements.
The application of the full Liouvillian is then implemented similar to a sparse matrix-vector multiplication.
Our code is available at \cite{thuemmler_2024_1}.

\subsection{Microscopic current}
The current operator is given by the commutator
\begin{equation} \label{eq:defCurrent}
    \hat{\mathbf{j}} = \mathrm{i} \left[ \hat{\mathbf{r}}, \hat{H}_0 \right]
    ,
\end{equation}
which intrinsically contains all sources of inter-, intra-, and anomalous current contributions \cite{wilhelm_semiconductor_2021,yue_2022}.
The intensity of the emitted radiation is obtained from the current $\mathbf{j} = \mathrm{Tr}(\hat{\mathbf{j}}\hat{\rho})$ via a Fourier transform (Larmor formula \cite{larmor_lxiii_1897}), i.e.
\begin{equation} \label{eq:defS}
    I(\omega) \sim  \omega^2 |\mathbf{j}(\omega)|^2
.
\end{equation}

\section{\label{sec:periodicModel}Real-space periodic model}
It is computationally prohibitive for the finite-size model to simulate QDs with sizes approaching the bulk limit.
In order to allow for a comparison with rt-TDDFT calculations for periodic models, we slightly adapt the model to obtain a periodic real-space model, which is computationally much easier.
It is based on the same real-space Wannier approach as the finite-size model, but tracks other coherences, i.e., off-diagonal elements of the density matrix, for the propagation.
Fig. \ref{fig:modelComparison} shows the differences in accounting for the coherences of the density matrix.
The finite-size model, see fig.~\ref{fig:modelComparison}a, allows only for coherences within the QD.
In the real-space periodic model, see fig.~\ref{fig:modelComparison}b, we additionally consider coherences, which arise from simultaneous shifts of both corresponding Wannier state sites. If $\rho_{\mathbf{xy},\mu\nu} \equiv \bra{\Psi_0} (\hat{e}_{\mathbf{y}\nu})^\dagger \hat{e}_{\mathbf{x}\mu} \ket{\Psi_0}$ is accounted for, then $\rho_{\mathbf{x}+\mathbf{s},\mathbf{y}+\mathbf{s},\mu\nu}$ is retained as well, where $\mathbf{s}$ is a super-cell lattice vector. 
Moreover, a coherence is considered only if its corresponding unit-cell centers are within a given distance, which we specify for each calculation individually.
The truncation of the coherences in the real-space periodic coherence distance is not identical to the treatment in the SBEs, see fig.~\ref{fig:modelComparison}c, where both sites may be independently shifted to other super cells. If $\rho_{\mathbf{xy}\mu\nu}$ is accounted for, than $\rho^{\mathbf{A}}_{\mathbf{x}+\mathbf{s},\mathbf{y},\mu\nu}$ is kept as well, where $\mathbf{s}$ is a super-cell lattice vector.
However, due to the translational symmetry, the corresponding density matrix elements during the propagation of the SBEs can be mapped to matrix elements based on sites of the same super cell by adding appropriate phase factors to the density matrix elements as indicated in fig.\ref{fig:modelComparison}c.
\begin{figure}
    \includegraphics[width=\linewidth]{modelComparison.pdf}
    \caption{Comparison of the considered spatial coherences within 
    a) the finite-size model,
    b) the real-space periodic model,
    and c) the SBEs
    along one dimension.
    The circles illustrate two different sites and their super cell periodic images.
    A solid (dashed) arrow indicates that the corresponding density matrix element is (not) accounted for during the propagation.
    The relations between the density matrix elements $\rho_{12}$ and their periodic counterparts  are depicted, where $A$ denotes the vector potential and $L$ is the length of the super cell.
}
    \label{fig:modelComparison}
\end{figure}
In appendix \ref{sec:sbe_equiv}, we show that our real-space periodic model in the limit of propagating all coherences is equivalent to the Fourier-transformed SBEs in the moving-frame Wannier basis \cite{silva_2019, yue_2022}.

\subsection{ Equations of motion }
We start the derivation for our real-space periodic model again from the Hamiltonian in eq.~\eqref{eq:HinLG}.
Contrary to the non-periodic case, we choose a moving-frame Wannier basis
\begin{equation} \label{eq:defPhasedWannier}
    \ket{w^{\mathbf{A}(t)}_{\mathbf{R}\alpha}} = \mathrm{e}^{\mathrm{i} \mathbf{A}(t) \cdot\mathbf{R}} \ket{\mathbf{R}\alpha}
\end{equation}
with the vector potential $\mathbf{A}(t)$.
For simplicity of notation, we drop the separate description of holes and electrons in this section and use electron operators only, i.e., $\hat{e}^{\mathbf{A}}_{\mathbf{R}\mu}$ is the annihilation operator corresponding to the Wannier function $\ket{w^{\mathbf{A}}_{\mathbf{R}\mu}}$.
The matrix elements of $\hat{H}_0$ and $\hat{\mathbf{r}} $ are given by
\begin{equation}
    H^{\mathbf{A}}_{\mathbf{xy}, \mu\nu} \equiv \bra{w^{\mathbf{A}}_{\mathbf{x}\mu}} \hat{H}_0 \ket{w^{\mathbf{A}}_{\mathbf{y}\nu}} = 
     \mathrm{e}^{\mathrm{i} (\mathbf{y}-\mathbf{x})\cdot\mathbf{A}} H_{\mathbf{y}-\mathbf{x},\mu\nu}
\end{equation}
and
\begin{equation}
    \mathbf{D}^{\mathbf{A}}_{\mathbf{xy}, \mu\nu} \equiv  \bra{w^{\mathbf{A}}_{\mathbf{x}\mu}} \hat{\mathbf{r}} \ket{w^{\mathbf{A}}_{\mathbf{y}\nu}} 
     = \mathrm{e}^{\mathrm{i} (\mathbf{y}-\mathbf{x})\cdot\mathbf{A}} \mathbf{D}_{\mathbf{y}-\mathbf{x},\mu\nu}
   + \delta_{\mathbf{x}\mathbf{y}}\delta_{\mu\nu} \mathbf{x}
\end{equation}
with $\mathbf{x}$ and $\mathbf{y}$ being lattice vectors, which define the unit cell where the corresponding Wannier functions are located.
The respective block of the density matrix is expressed as
\begin{equation} \label{eq:defPhasedRho}
    \rho^{\mathbf{A}}_{\mathbf{x}\mathbf{y},\mu\nu} = \bra{\Psi_0} (\hat{e}^\mathbf{A}_{\mathbf{y}\nu})^\dagger \hat{e}^\mathbf{A}_{\mathbf{x}\mu} \ket{\Psi_0}
\end{equation}

From now on, we omit all Wannier function indices and define the matrices
\begin{equation} \label{eq:defHAE}
    H^{\mathbf{A},\mathbf{E}}_{\mathbf{xy}} \equiv  H^{\mathbf{A}}_{\mathbf{xy}} + \mathbf{E} \cdot\mathbf{D}^{\mathbf{A}}_{\mathbf{xy}}
    .
\end{equation}
The time derivative of the density matrix obeys the von-Neumann equation
\begin{equation} \begin{aligned}
    \mathrm{i} \absDeriv{t}{\rho^{\mathbf{A}}_{\mathbf{xy}} } &= 
        \sum\limits_{\mathbf{s}}
                    (\rho^{\mathbf{A}}_\mathbf{xs} H^{\mathbf{A},\mathbf{E}}_{\mathbf{sy}}
                    -  H^{\mathbf{A},\mathbf{E}}_\mathbf{xs}  \rho^{\mathbf{A}}_\mathbf{sy})
            + \mathbf{E}(\mathbf{x}-\mathbf{y}) \rho^{\mathbf{A}}_{\mathbf{xy}}
            \\
            &=
        \sum\limits_{\mathbf{s}}
                    (\rho^{\mathbf{A}}_\mathbf{xs} H^{\mathbf{A},\mathbf{E}}_{\mathbf{0},\mathbf{y}-\mathbf{s}}
                    -  H^{\mathbf{A},\mathbf{E}}_{\mathbf{0},\mathbf{s}-\mathbf{x}}  \rho^{\mathbf{A}}_\mathbf{sy})
,
\end{aligned}\end{equation}
where the sum is taken over all lattice vectors $\mathbf{s}$.
The ground state and the evolution of the density matrix has a translational invariant form, i.e., $\rho^{\mathbf{A}}_{\mathbf{xy}} = \rho^{\mathbf{A}}_{\mathbf{x}+\mathbf{s},\mathbf{y}+\mathbf{s}}$ for an arbitrary lattice vector $\mathbf{s}$.
Therefore, the definitions
$\rho^{\mathbf{A}}_{\mathbf{x}} \equiv \rho^{\mathbf{A}}_{\mathbf{0x}}$ and $ H^{\mathbf{A},\mathbf{E}}_{\mathbf{x}} = H^{\mathbf{A},\mathbf{E}}_{\mathbf{0},\mathbf{x}}$ lead us to
\begin{equation} \label{eq:prop_periodic}
    \mathrm{i} \absDeriv{t}{\rho^{\mathbf{A}}_{\mathbf{x}} } =
        \sum\limits_{\mathbf{s}} \rho^{\mathbf{A}}_{\mathbf{s}} H^{\mathbf{A},\mathbf{E}}_{\mathbf{x}-\mathbf{s}} - H^{\mathbf{A},\mathbf{E}}_{\mathbf{s}} \rho^{\mathbf{A}}_{\mathbf{x}-\mathbf{s}}
.
\end{equation}
We note, that $\rho_{\mathbf{x}}$ describes the coherences between those Wannier states, whose corresponding unit cell center positions differ by $\mathbf{x}$.
\subsection{Microscopic current}
The current,  see eq.~\eqref{eq:defCurrent}, is obtained from the density matrix as
\begin{equation} \begin{aligned} \label{eq:current_periodicDeriv}
    \mathbf{j} &= \mathrm{i}\,\mathrm{Tr}\sum\limits_{\mathbf{axy}} \rho_{\mathbf{xy}}^{\mathbf{A}}
        (\mathbf{D}^{\mathbf{A}}_\mathbf{xa} H^{\mathbf{A}}_\mathbf{ay} - H^{\mathbf{A}}_\mathbf{xa} \mathbf{D}^{\mathbf{A}}_\mathbf{ay})
              \\
              &= 
                  \mathrm{i}\,\mathrm{Tr} \sum\limits_{\mathbf{xy}} \rho_{\mathbf{xy}}^{\mathbf{A}} \biggl[
                  \sum\limits_{\mathbf{a}} 
        (\mathbf{D}^{\mathbf{A}}_{\mathbf{a}-\mathbf{x}} H^{\mathbf{A}}_{\mathbf{y}-\mathbf{a}} 
                - \mathbf{H}^{\mathbf{A}}_{\mathbf{a}-\mathbf{x}} D^{\mathbf{A}}_{\mathbf{y}-\mathbf{a}})
             \\  &\quad\quad\quad\quad\quad\quad\quad
                + (\mathbf{x}-\mathbf{y})H^{\mathbf{A}}_{\mathbf{y}-\mathbf{x}} \biggr]
.
\end{aligned} \end{equation}
By invoking the translational invariance of the density matrix elements, we derive the current per unit cell to be
\begin{equation}\label{eq:current_periodic}
    \mathbf{j} = \mathrm{i}\,\mathrm{Tr} \sum\limits_{\mathbf{s}} \rho_\mathbf{s}^{\mathbf{A}}\left[\sum\limits_{\mathbf{a}} 
        \mathbf{D}^{\mathbf{A}}_{\mathbf{a}} H^{\mathbf{A}}_{\mathbf{s}-\mathbf{a}} 
                - H^{\mathbf{A}}_{\mathbf{a}} \mathbf{D}^{\mathbf{A}}_{\mathbf{s}-\mathbf{a}}
                - \mathbf{s} H^{\mathbf{A}}_{\mathbf{s}} \right]
.
\end{equation}
We note, that the trace in the equations above is taken over the Wannier indices.
Equation \eqref{eq:current_periodic} enables us to calculate the emitted radiation from the Fourier transform of $\mathbf{j}$ as defined in eq.~\eqref{eq:defS}.

\subsection{Numerical Propagation}
The time derivative of the density matrix $\rho_{\mathbf{x}}$ in eq.~\eqref{eq:prop_periodic} shows, that finite coherences for every lattice vector $\mathbf{x}$ may build up.
For the numerical treatment, we have to restrict the number of allowed coherence distance vectors $\mathbf{x}$.
We call the corresponding finite set, coherence region and neglect the coherences of the density matrix outside of this region in the evaluation of eq.~\eqref{eq:prop_periodic}.
For all calculations presented in this work, we have chosen the coherence region as large as computationally feasible, typically including up to $\sim 10^3$ unit cells.
We define the coherence regions later for each calculation separately.
Lastly, we define the initial state to have completely filled valence and empty conduction bands. 
We developed our code \cite{thuemmler_2024_2} for CPUs and GPUs, where we integrate the equation of motion using the Runge-Kutta 4/5 algorithm implemented in the C++ boost odeint library \cite{ahnert_2011} in combination with vexcl \cite{demidov_2013}.

\section{\label{sec:numeric}Simulation results}
We benchmark our real-space periodic model for bulk silicon and CdSe (wurtzite) within the Wannier framework.
We calculate the optical conductivity and compare it with the results of the Greenwood-Kubo formula \cite{greenwood_1958,pizzi_2020}.
We then contrast our real-space periodic model to computationally expensive rt-TDDFT calculations of HHG for bulk silicon.
After this we consider the finite-size model.
First, we simulate the optical response of a $\mathrm{Cd}_{11}\mathrm{Se}_{11}$ cluster, see fig.~\ref{fig:geometries}a for its geometry using rt-TDDFT and compare it to our model.
Second, we demonstrate, that the finite-size model reproduces the experimentally observed drop in HHG yield for small dots.
Lastly, we investigate the ellipticity dependence of the HHG yield for a CdSe QD with a diameter of 2.8\,nm, see fig.~\ref{fig:geometries}b for its geometry.

\subsection{Parameterization}
We set up two tight-binding models, one for silicon and the other for CdSe in the wurtzite phase.
For both, we performed DFT calculations without spin-orbit coupling via the plane-wave code of Quantum Espresso (QE) 7.2 \cite{giannozzi_2009,giannozzi_2017,giannozzi_2020} on a $9\times9\times9$ Monkhorst-Pack grid \cite{monkhors_1976}.
For silicon we employed the pseudopotential from ref.~\cite{von_barth_1980} within the local density approximation (LDA).
For CdSe, we used norm-conserving potentials \cite{dal_corso_1996,schlipf_2015} parametrized for the PBE functional \cite{perdew_1996} and shifted the band gap energy to the experimental value of $1.75\,\mathrm{eV}$ \cite{ninomiya_1995}.

\begin{figure}
    \includegraphics[width=\linewidth]{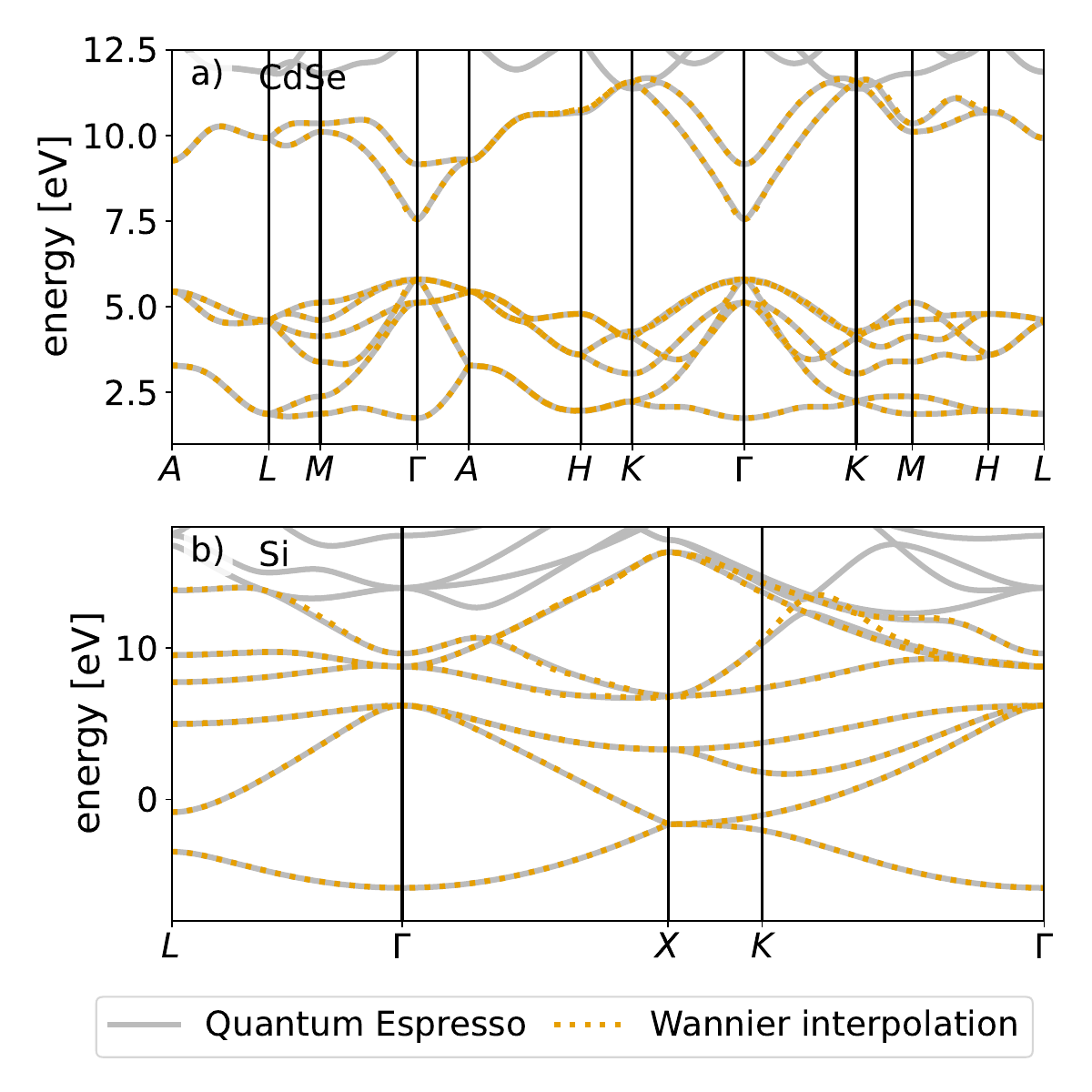}
    \caption{DFT band structures calculated via Quantum Espresso together with their Wannier interpolations for a) CdSe (wurtzite) and b) Si.}
    \label{fig:bandStructure}
\end{figure}

Based on the DFT calculations, we performed a Wannierization using Wannier90 \cite{pizzi_2020}. 
We separately Wannierized the conduction and valence bands to use them in our finite-size model.
The silicon model is setup from four Wannier functions for the conduction and valence bands, each.
For CdSe, we calculated six Wannier functions to represent the valence bands and two for the conduction bands.
The corresponding band structures and the Wannier interpolated ones are shown in fig.~\ref{fig:bandStructure}.
For our model, we also calculated the transition dipole moments between the different Wannier functions.
We note that the dipole matrix elements, see eq.~\eqref{eq:defWannierR}, provided by Wannier90 do not obey $\mathbf{D}_{\mathbf{R}, \mu\nu} = \mathbf{D}^{*}_{-\mathbf{R}, \nu\mu}$.
As our propagation scheme relies on the hermiticity of the Hamiltonian, we first symmetrized those matrix elements.

\subsection{Linear optical properties -- absorption}
\begin{figure} 
    \includegraphics[width=\linewidth]{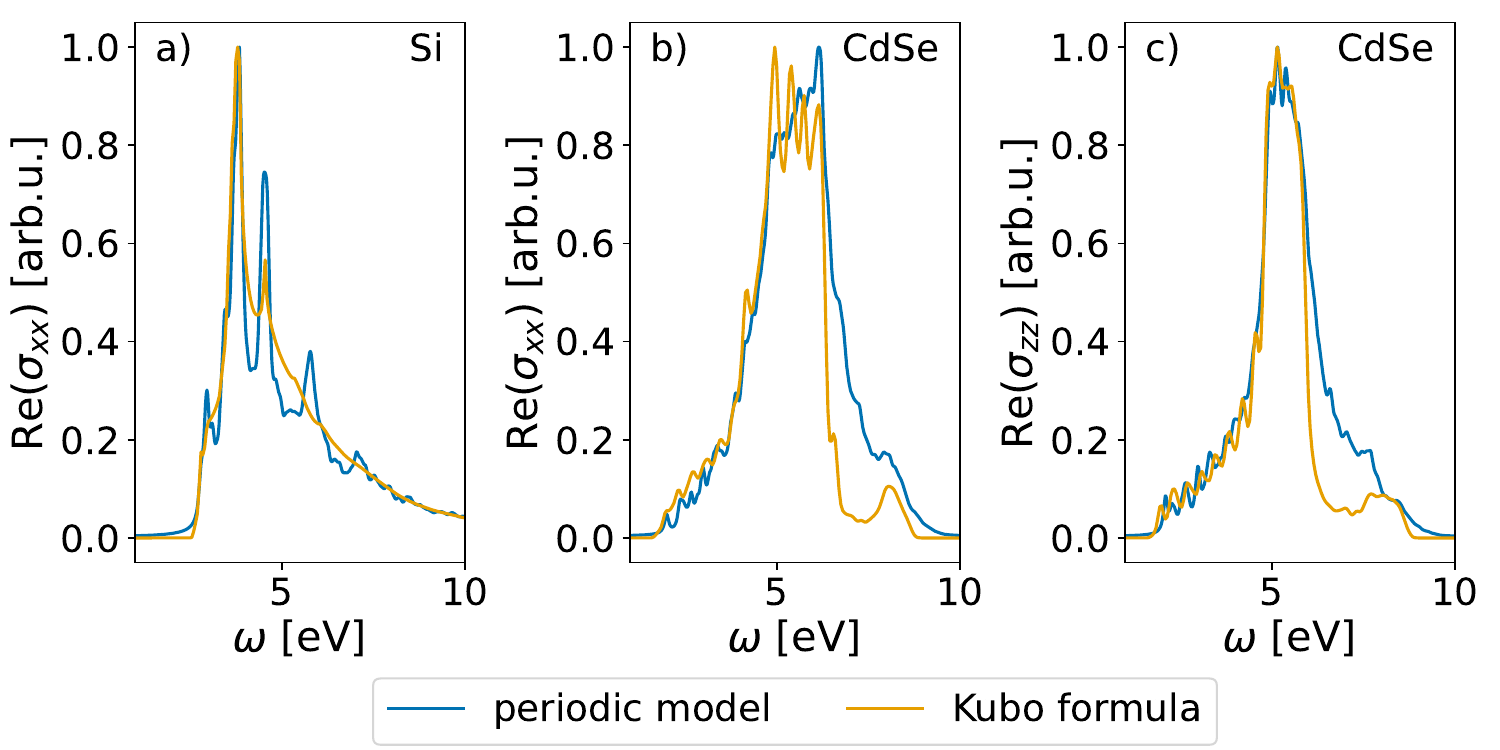}
    \caption{\label{fig:opticalConductivity} Selected tensor components of the normalized optical conductivity $\sigma(\omega)$ calculated via the real-space periodic model and via Kubo formula for a) Si and  CdSe (wurtzite) along the b) a-axis and c) c-axis. }
\end{figure}

To approve our model for low field strengths, we calculated the optical conductivity tensor $\sigma$, which is defined via $\mathbf{j}(\omega) = \sigma(\omega) \mathbf{E}(\omega)$,  within our real-space periodic tight-binding approach.
For that purpose, we propagated the density matrix in an electric field of Gaussian shape, i.e.,
\begin{equation}
    \mathbf{E}(t) = \mathbf{E}_0\mathrm{e}^{-\frac{t^2}{2t_\mathrm{W}^2}}
\end{equation}
with amplitude $|\mathbf{E}_0| = 1\mathrm{V}/\mu\mathrm{m}$, and  $t_\mathrm{W}=\sqrt{8\ln2}\,t_\mathrm{FWHM}$ with $t_{\mathrm{FHWM}} = 0.1\,\mathrm{fs}$ from $t=-1\,\mathrm{fs}$ to $t=100\,\mathrm{fs}$.
The resulting current $\mathbf{j}(t)$ is damped with $\exp({-t/t_{\mathrm{damp}}})$ with $t_\mathrm{damp} = 10\,\mathrm{fs}$ to obtain the real part of the conductivity via Fourier transformation.
We ran the calculations for three linearly polarized pulses along the $\mathbf{x}$, $\mathbf{y}$ and $\mathbf{z}$ direction to extract $\sigma_{xx}$, $\sigma_{yy}$, and $\sigma_{zz}$, respectively.
For comparison, we employed the postw90 program \cite{pizzi_2020} on a $125\times125\times125$ $\mathbf{k}$-grid, and calculated the optical conductivity using the Kubo formula \cite{greenwood_1958} directly.
Fig. \ref{fig:opticalConductivity} depicts the frequency-dependent optical conductivity of Si and CdSe for selected tensor components calculated by both methods.

For silicon, we took (for computational reasons) only those Hamiltonian and dipole matrix elements into account for which the centers of the corresponding units cells have a distance of less than 2.2\,nm (approximately 6 unit cells in each direction) and ensured that this truncation did not alter the Hamiltonian significantly.
We used a coherence region during the propagation of the density matrix, defined by a maximum distance of 2.5\,nm and 2.2\,nm along the polarization and all orthogonal directions, respectively.
Because of the cubic symmetry, we restrict our discussion to $\sigma_{xx}$, see fig.~\ref{fig:opticalConductivity}a.
Our model reproduces the sharp onset of the absorption at the direct DFT-band gap of $2.5\,\mathrm{eV}$.
The positions of the main peaks agree very well with the reference Kubo-conductivity for silicon.
There are some differences in the amplitudes and widths of those peaks, which we attribute to the different line broadening and the much smaller number of real space grid points ($\sim 10^2$) compared to the fine $\mathbf{k}$-mesh ($\sim 10^6$ points) on which the Kubo formula was evaluated.

For CdSe, we truncated the Hamiltonian matrix elements at a unit cell distance of 2.3\,nm (5 and 3 unit cells along a-axis and c-axis direction, respectively).
We propagated the coherences of the density matrix for all unit cells with a maximum distance of 12.5\,nm and 2.8\,nm along polarization and all orthogonal directions, respectively.
Fig.~\ref{fig:opticalConductivity}a and b displays $\sigma_{xx}$ and $\sigma_{zz}$ referring to the a-axis and c-axis of the crystal, respectively.
The absorption spectrum calculated via the periodic real-space model agrees very well with the Kubo formula below 5\,eV.
For energies higher than 6\,eV the trend is correct, but our model overestimates the absorption.
We attribute this to the limited number of bands in our model.
Both examples show that our model correctly captures the linear response to an electric field within the Wannier framework.

\subsection{Nonlinear optical properties} \label{sec:rttddft}
In this section, we contrast both models to rt-TDDFT calculations of HHG for linearly polarized pulses.
We simulated the HHG for silicon with our periodic model and compare it with Octopus \cite{tancogne-dejean_2020} for a linearly polarized pulse, which is described by the vector potential
\begin{equation} \label{eq:sin2Pulse}
    \mathbf{A}(t) = \frac{\mathbf{E}_0}{\omega} \sin^2\left(\frac{\pi t}{2t_{\mathrm{FWHM}}}\right) \sin(\omega t)
\end{equation}
from $t=-t_{\mathrm{FWHM}}$ to $t=t_{\mathrm{FWHM}}$.
It has a full-width at half-maximum (FWHM) of $t_{\mathrm{FWHM}}=30\,\mathrm{fs}$, a maximum field strength $|\mathbf{E}_0|= 1\,\mathrm{V}/\mathrm{nm}$, and a wavelength $\lambda = \frac{2\pi c}{\omega}= 2\,\mu\mathrm{m}$, where $c$ is the speed of light.
For the Octopus HHG calculation \cite{tancogne-dejean_2017_1}, we employed a $28\times28\times28$  $\mathbf{k}$-grid, a real-space discretization of $0.344\,\mathrm{a.u.}$ along each lattice vector direction, and a time step of $0.02\,\mathrm{a.u}$.
In the periodic model, we truncated the Hamiltonian in real-space as described above and took coherences of the density matrix up to a maximum distance of 2.8\,nm (approximately 7 unit cells in each direction) into account.
The rt-TDDFT calculations took a few days, whereas our approach required at most 20 minutes to simulate a single pulse.
\begin{figure}
        \includegraphics[width=\linewidth]{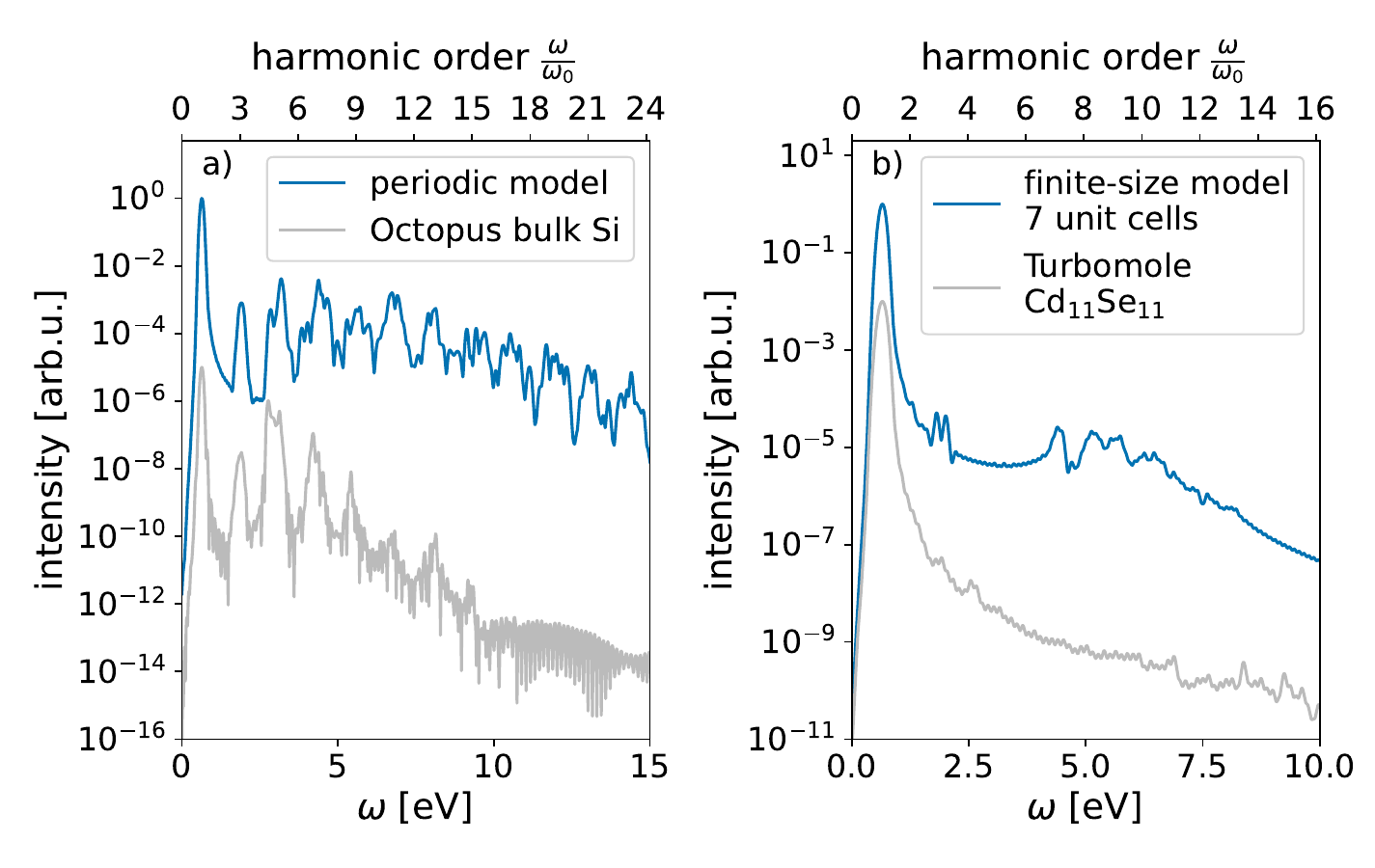}
    \caption{\label{fig:refHHG}Emission spectra of our models compared with rt-TDDFT calculations for
             a) periodic silicon (real-space periodic model) and
             b) a $\mathrm{Cd}_{11}\mathrm{Se}_{11}$ cluster (finite-size model).
             The spectra are normalized to their maxima and shifted with respect to each other to improve visibility.
            }
\end{figure}
Both simulation approaches show qualitatively good agreement, see fig. \ref{fig:refHHG}a.
In comparison with the rt-TDDFT calculation, our model predicts a stronger intensity of the higher-harmonics (even above $10$ eV).
Note that the intensity of the bulk HHG sensitively depends on dephasing \cite{vampa_2014}, which can be easily incorporated in tight-binding models.
However, we neglected dephasing effects here as their proper modeling, especially for finite structures, is out of the scope of this study.

Next, we simulated HHG for the biggest CdSe cluster for which we could run an rt-TDDFT simulation for a single pulse within some days of calculation time using the same pulse as above, see eq.~\eqref{eq:sin2Pulse}.
One pulse for these small QDs is modeled by our finite-size model within a few seconds. 
We performed an rt-TDDFT calculation for the $\mathrm{Cd}_{11}\mathrm{Se}_{11}$ quantum dot (diameter $<1$\,nm) using the riper module \cite{muller_2020} of turbomole \cite{franzke_2023}.
The geometry of the QD was obtained by optimizing the atomic positions of a spherical cut-out from a CdSe (wurtzite) bulk structure.
We employed xtb \cite{bannwarth_2021} with the GFN1 \cite{grimme_2017} force field for an initial geometry optimization and then used the default optimizer of turbomole to obtain the ground state geometry, see fig.~\ref{fig:geometries}a, employing the PBE functional \cite{perdew_1996}.
For the time propagation, we used a time step of $0.02\,\mathrm{a.u.}$ and the TZVP-basis \cite{weigend_2005} for Cd and Se.
Additionally, we utilized the resolution of identity technique with the universal basis set \cite{weigend_2006}.
Note, that the small cluster as displayed in fig.\ref{fig:geometries}a is by far not spherical and its optical response is highly directional.
Hence, a reasonable agreement with experimental results requires averaging over many orientations, which is beyond our numerical capacities for rt-TDDFT calculations.
The finite-size calculations are based on a QD model with 7 unit cells (dot diameter $1\,\mathrm{nm}$), forming a hexagon like structure and mimicking a $\mathrm{Cd}_{14}\mathrm{Se}_{14}$ cluster.
To suppress any directional dependence of this small non-spherical model system, we averaged the HHG yield over 100 random orientations of the QD with respect to the pulse polarization.
Fig. \ref{fig:refHHG}a shows the obtained HHG yield. 
For our finite-size model as well as the rt-TDDFT calculation, clearly no higher harmonics are generated, which is in accordance with the experiments.
To compare the rt-TDDFT and our tight-binding approach for a specific CdSe cluster, we could refine the model parametrization by employing a ground state DFT calculation for that specific geometry.

\subsection{Quantum dot-size and ellipticity dependence of the HHG yield} \label{sec:sizeDep}
After the validation of our model, we demonstrate its capabilities to simulate ensemble averages of HHG in CdSe QDs with diameters of up to $3.3\,\mathrm{nm}$ ($\sim 10^3$ Wannier functions).
This is possible as the simulation of a single pulse takes, even for the largest QD-sizes, at most 30 minutes on an Nvidia A100 graphics card.

\begin{figure*}
    \includegraphics[width=\textwidth]{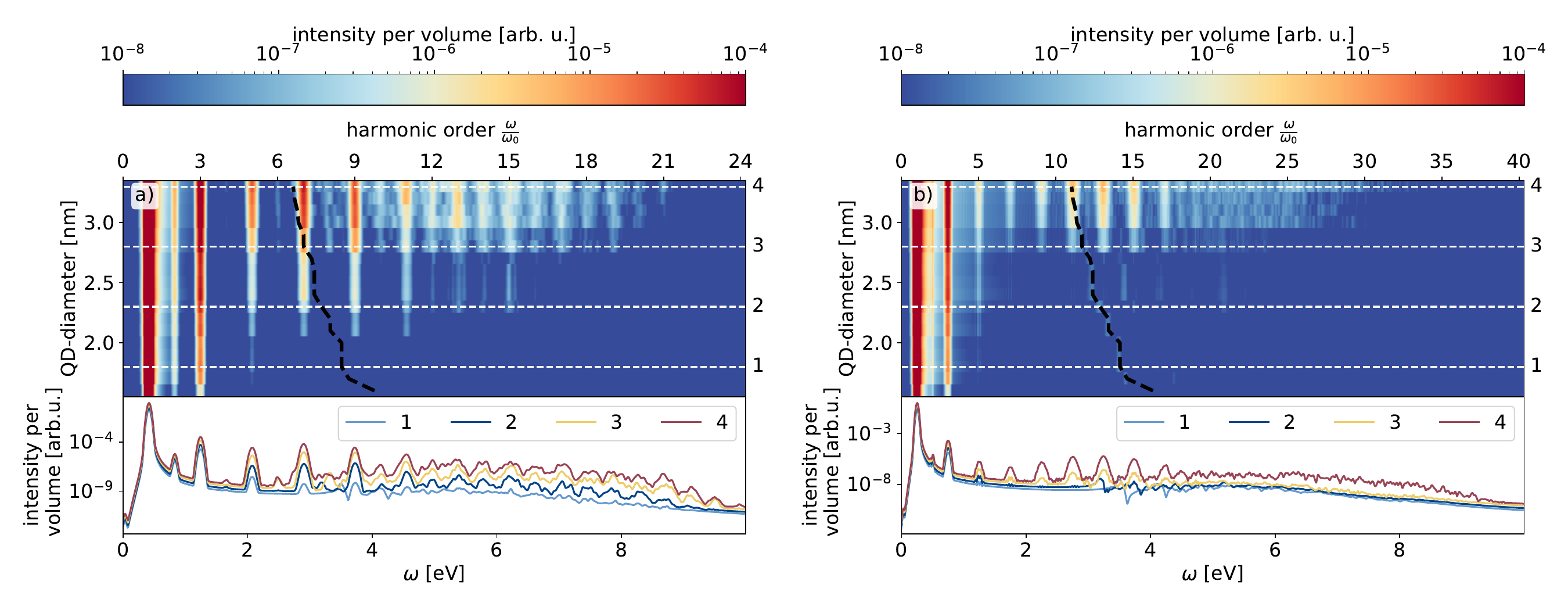}
    \caption{Averaged spectral yield as function of the CdSe QD-size for linear polarized pulses with a) $\lambda=3\,\mu\mathrm{m}$ and b) $\lambda=5\,\mu\mathrm{m}$.
             The intensity is scaled per dot volume and normalized to its maximum.
             The black dashed lines indicate the QD-size dependent band gap obtained from the position of the energetically lowest peak in the linear absorption spectra.
             The lower panels depict the spectra for selected dot sizes that are indicated by the horizontally oriented white dashed lines.
             }
     \label{fig:qdSize}
\end{figure*}

First, we demonstrate that our model successfully reproduces the QD-size dependence of the HHG yield as observed experimentally \cite{gopalakrishna_2023,nakagawa_2022}.
HHG was simulated for a linearly polarized light pulse with a maximum field strength of $E_0$ = 1\,V/nm (= 0.3\,TW/cm$^2$) and a FWHM of 100\,fs using the same shape as defined in eq.~\eqref{eq:sin2Pulse}.
To allow for orientation-independent conclusions, we averaged all spectra over the same randomly selected 100 dot orientations.
Fig.~\ref{fig:qdSize} depicts the calculated spectra for different dot sizes and for driving wavelengths of $\lambda = 3\,\mu\mathrm{m}$ and $\lambda = 5\,\mu\mathrm{m}$.
For both wavelengths, no above band gap harmonics are generated for dots smaller than 2\,nm.
For slightly larger dot sizes, the higher harmonics emerge for $\lambda=3\,\mu\mathrm{m}$ much faster and stronger than for $\lambda=5\,\mu\mathrm{m}$.
Lastly, the yield of the higher harmonics increases only slightly for dot diameters from 3\,nm to 3.3\,nm.
Recall that all of these trends were observed experimentally \cite{gopalakrishna_2023}.
As before, the suppression of the higher harmonics can be explained with the three-step model for HHG in solids \cite{vampa_2014,ikemachi_trajectory_2017}, which we assume to be applicable to these QDs as well.
The first step is the excitation of electron-hole pairs, which is not affected by the dot geometry.
However, the second step, the acceleration of the carriers is impacted by the finite size of the dots.
Moreover, the dot boundaries induce dephasing and scattering of the electrons and holes.
In particular, the latter process is more pronounced for longer wavelengths due to higher particle momenta (increased maximum vector potential).
Consequently, the third step, the recombination of electron-hole pairs creating a inter-band contribution to the higher-harmonic radiation, is suppressed more strongly for longer driving wavelengths in the small dots.

\begin{figure}
    \includegraphics[width=\linewidth]{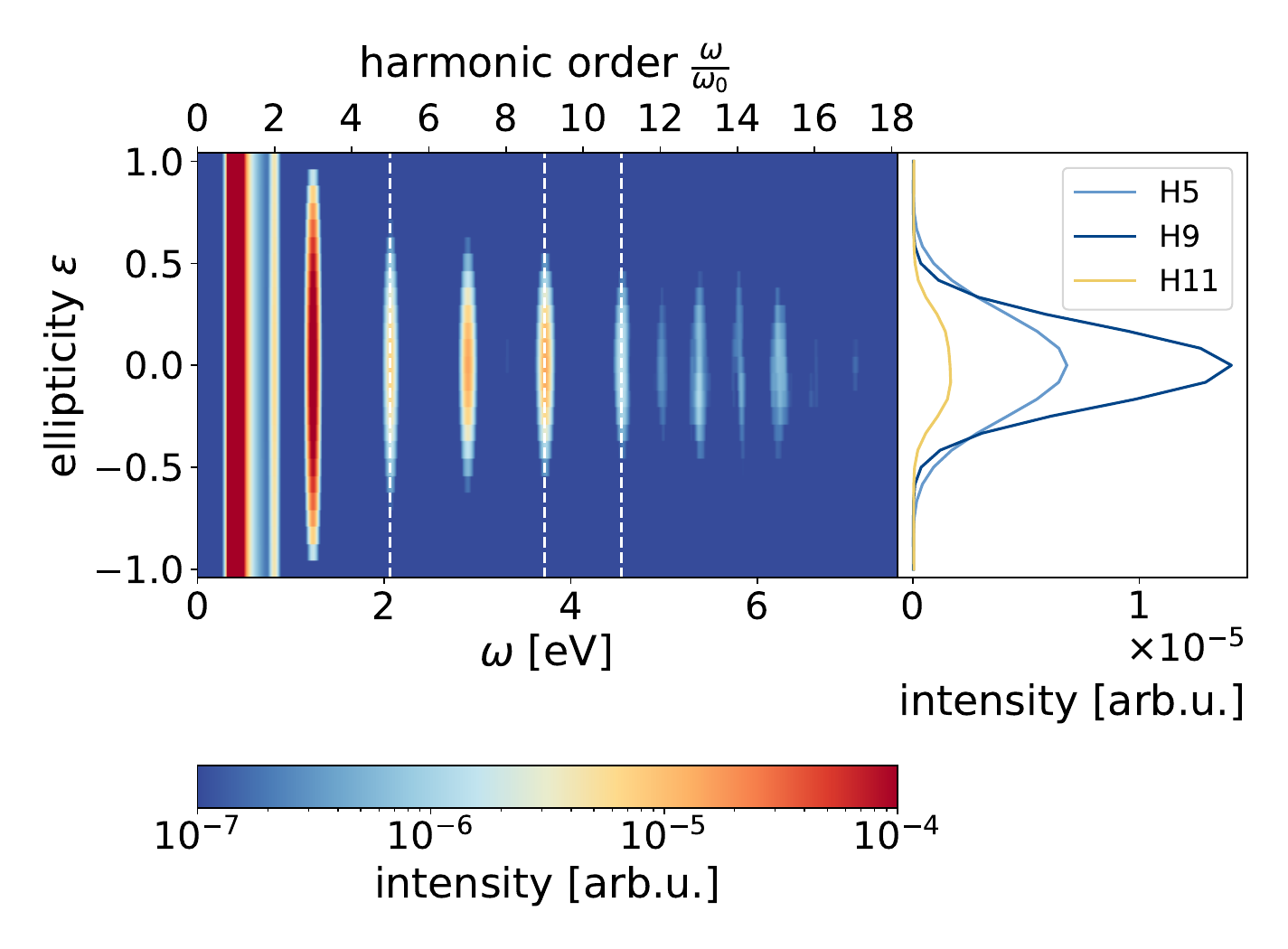}
    \caption{Spectral yield as function of the ellipticity $\epsilon$ of the driving pulse for a CdSe QD of diameter 2.8\,nm and $\lambda=3\,\mu\mathrm{m}$.
    The intensity is normalized to its maximum.}
    \label{fig:qdEl}
\end{figure}

\begin{figure}
    \includegraphics[width=\linewidth]{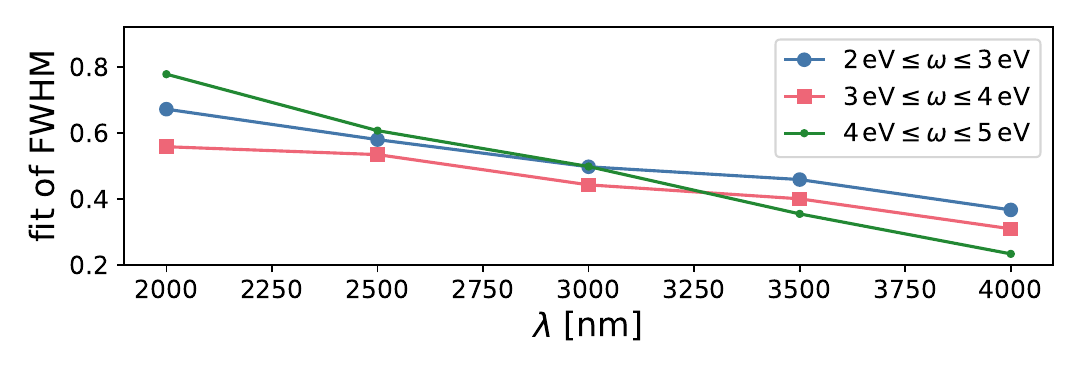}
    \caption{FWHM of a Gaussian-fit of the ellipticity dependent yield as function of the driving wavelength integrated over the indicated energy ranges.
    The CdSe QD-diameter is 2.8\,nm.}
    \label{fig:elFreqScan}
\end{figure}

Finally, we investigate the wavelength and ellipticity dependence of the above band gap harmonics.
We use the same pulse shapes and orientations as before, but allow for different ellipticities.
An ellipticity of $\epsilon = E_{0x}/E_{0y} = 1\, (-1)$ corresponds to a right (left) circular polarized pulse.
Fig. \ref{fig:qdEl} depicts the HHG yield of a QD with a diameter of 2.8\,nm for different ellipticities of the driving pulse with $\lambda =3\,\mu{\mathrm{m}}$.
We refer for its geometry to fig.~\ref{fig:geometries}b.
Due to the broken inversion symmetry of the system, even-order harmonics are generated in each dot, but average out for large ensembles with random orientations.
Indeed, some even order harmonics are still visible in fig.~\ref{fig:qdEl}, but their intensity is much lower than that of the odd-order harmonics.
They are expected to disappear completely for an average that involves even larger ensembles.
With increasing ellipticity, the higher-order harmonics are suppressed strongly, similar to the bulk and in agreement with experiments \cite{gopalakrishna_2023}.
The integrated intensity for each harmonic as function of the ellipticity $\epsilon$ follows a Gaussian-like dependence.
Fig.~\ref{fig:elFreqScan} shows the Gaussian-fitted FWHM of the ellipticity dependence integrated over fixed energy ranges for different driving wavelengths $\lambda$ of the same QD (diameter $2.8\,\mathrm{nm}$).
The lowest indicated energy range (2\,eV to 3\,eV) covers the harmonics up to the band gap ($\approx 3\,\mathrm{eV}$, see dashed black line in fig.\ref{fig:qdSize}).
It behaves qualitatively identically to the next energy range above the band gap (3\,eV to 4\,eV): The FWHM drops with increasing driving pulse-wavelength, which was seen in bulk ZnO \cite{ghimire_2011,liu_2016}, with very similar values of the FWHM (ranging from 0.25 to 0.5).
The highest energy range (4\,eV to 5\,eV) has a much steeper reduction of the FWHM with increasing wavelength than the others.
This coincides with the size-dependent suppression in this energy range for the longer driving wavelengths, which is depicted by the dashed line 3 in fig.~\ref{fig:qdSize}a.
Therefore, we conclude that the ellipticity dependence of the higher-order harmonics is a combination of the bulk behavior and boundary effects.

\section{\label{sec:conclusion}Conclusion}
We presented a three-dimensional real-space tight-binding model to simulate HHG in medium-sized spatially confined structures and applied it to CdSe QDs.
This finite-size model is based on the representation of the Hamiltonian and the dipole matrices in terms of Wannier functions, which can be obtained from ab-initio calculations.
To benchmark the periodic bulk limit of our model, we introduced a real-space periodic model by adapting the coherences accounted for during the propagation of the the density matrix.
This model was used for the calculation of absorption and HHG spectra, which were successfully compared to respective rt-TDDFT calculations.
Our efficient GPU parallelized implementation of the finite-size model allows to propagate the density matrix for QDs of  diameters of up to $3.3\,\mathrm{nm}$ ($\sim 10^3$ Wannier functions), which is intractable within rt-TDDFT calculations.
To showcase the capabilities of our finite-size model, we calculated ensemble averages of HHG spectra over different QD orientations in the medium-size regime, to reproduce the experimentally observed QD-size dependence of the HHG yield for linear polarization \cite{nakagawa_2022,gopalakrishna_2023}.
Additionally, we simulated the response to different ellipticities of the light pulses, yielding a similar behavior to that of bulk materials, but with clear indications of finite-size effects.
Lastly, we emphasize that our model is based on ab-initio parameterization and therefore transferrable to other nanostructures as well and bridges the gap between the description of HHG for atoms or molecules and solids.

\section{Acknowledgment}
The project was funded by the Deutsche Forschungsgemeinschaft (DFG, German Research Foundation) – Project-ID 398816777 – SFB 1375, A1.

\bibliography{refs}

\begin{thebibliography}{52}%
\makeatletter
\providecommand \@ifxundefined [1]{%
 \@ifx{#1\undefined}
}%
\providecommand \@ifnum [1]{%
 \ifnum #1\expandafter \@firstoftwo
 \else \expandafter \@secondoftwo
 \fi
}%
\providecommand \@ifx [1]{%
 \ifx #1\expandafter \@firstoftwo
 \else \expandafter \@secondoftwo
 \fi
}%
\providecommand \natexlab [1]{#1}%
\providecommand \enquote  [1]{``#1''}%
\providecommand \bibnamefont  [1]{#1}%
\providecommand \bibfnamefont [1]{#1}%
\providecommand \citenamefont [1]{#1}%
\providecommand \href@noop [0]{\@secondoftwo}%
\providecommand \href [0]{\begingroup \@sanitize@url \@href}%
\providecommand \@href[1]{\@@startlink{#1}\@@href}%
\providecommand \@@href[1]{\endgroup#1\@@endlink}%
\providecommand \@sanitize@url [0]{\catcode `\\12\catcode `\$12\catcode
  `\&12\catcode `\#12\catcode `\^12\catcode `\_12\catcode `\%12\relax}%
\providecommand \@@startlink[1]{}%
\providecommand \@@endlink[0]{}%
\providecommand \url  [0]{\begingroup\@sanitize@url \@url }%
\providecommand \@url [1]{\endgroup\@href {#1}{\urlprefix }}%
\providecommand \urlprefix  [0]{URL }%
\providecommand \Eprint [0]{\href }%
\providecommand \doibase [0]{http://dx.doi.org/}%
\providecommand \selectlanguage [0]{\@gobble}%
\providecommand \bibinfo  [0]{\@secondoftwo}%
\providecommand \bibfield  [0]{\@secondoftwo}%
\providecommand \translation [1]{[#1]}%
\providecommand \BibitemOpen [0]{}%
\providecommand \bibitemStop [0]{}%
\providecommand \bibitemNoStop [0]{.\EOS\space}%
\providecommand \EOS [0]{\spacefactor3000\relax}%
\providecommand \BibitemShut  [1]{\csname bibitem#1\endcsname}%
\let\auto@bib@innerbib\@empty
\bibitem [{\citenamefont {Ferray}\ \emph {et~al.}(1988)\citenamefont {Ferray},
  \citenamefont {L'Huillier}, \citenamefont {Li}, \citenamefont {Lompre},
  \citenamefont {Mainfray},\ and\ \citenamefont {Manus}}]{ferray_1988}%
  \BibitemOpen
  \bibfield  {author} {\bibinfo {author} {\bibfnamefont {M.}~\bibnamefont
  {Ferray}}, \bibinfo {author} {\bibfnamefont {A.}~\bibnamefont {L'Huillier}},
  \bibinfo {author} {\bibfnamefont {X.~F.}\ \bibnamefont {Li}}, \bibinfo
  {author} {\bibfnamefont {L.~A.}\ \bibnamefont {Lompre}}, \bibinfo {author}
  {\bibfnamefont {G.}~\bibnamefont {Mainfray}}, \ and\ \bibinfo {author}
  {\bibfnamefont {C.}~\bibnamefont {Manus}},\ }\bibfield  {title} {\enquote
  {\bibinfo {title} {Multiple-harmonic conversion of 1064 nm radiation in rare
  gases},}\ }\href {\doibase 10.1088/0953-4075/21/3/001} {\bibfield  {journal}
  {\bibinfo  {journal} {Journal of Physics B: Atomic, Molecular and Optical
  Physics}\ }\textbf {\bibinfo {volume} {21}},\ \bibinfo {pages} {L31--L35}
  (\bibinfo {year} {1988})}\BibitemShut {NoStop}%
\bibitem [{\citenamefont {Ghimire}\ \emph {et~al.}(2011)\citenamefont
  {Ghimire}, \citenamefont {DiChiara}, \citenamefont {Sistrunk}, \citenamefont
  {Agostini}, \citenamefont {DiMauro},\ and\ \citenamefont
  {Reis}}]{ghimire_2011}%
  \BibitemOpen
  \bibfield  {author} {\bibinfo {author} {\bibfnamefont {S.}~\bibnamefont
  {Ghimire}}, \bibinfo {author} {\bibfnamefont {A.~D.}\ \bibnamefont
  {DiChiara}}, \bibinfo {author} {\bibfnamefont {E.}~\bibnamefont {Sistrunk}},
  \bibinfo {author} {\bibfnamefont {P.}~\bibnamefont {Agostini}}, \bibinfo
  {author} {\bibfnamefont {L.~F.}\ \bibnamefont {DiMauro}}, \ and\ \bibinfo
  {author} {\bibfnamefont {D.~A.}\ \bibnamefont {Reis}},\ }\bibfield  {title}
  {\enquote {\bibinfo {title} {Observation of high-order harmonic generation in
  a bulk crystal},}\ }\href {\doibase 10.1038/nphys1847} {\bibfield  {journal}
  {\bibinfo  {journal} {Nature Physics}\ }\textbf {\bibinfo {volume} {7}},\
  \bibinfo {pages} {138--141} (\bibinfo {year} {2011})}\BibitemShut {NoStop}%
\bibitem [{\citenamefont {Li}\ \emph {et~al.}(2023)\citenamefont {Li},
  \citenamefont {Saleh}, \citenamefont {Sharma}, \citenamefont {Hünecke},
  \citenamefont {Sierka}, \citenamefont {Neuhaus}, \citenamefont {Hedewig},
  \citenamefont {Bergues}, \citenamefont {Alharbi}, \citenamefont {ALQahtani},
  \citenamefont {Azzeer}, \citenamefont {Gräfe}, \citenamefont {Kling},
  \citenamefont {Alharbi},\ and\ \citenamefont {Wang}}]{li_resonance_2023}%
  \BibitemOpen
  \bibfield  {author} {\bibinfo {author} {\bibfnamefont {W.}~\bibnamefont
  {Li}}, \bibinfo {author} {\bibfnamefont {A.}~\bibnamefont {Saleh}}, \bibinfo
  {author} {\bibfnamefont {M.}~\bibnamefont {Sharma}}, \bibinfo {author}
  {\bibfnamefont {C.}~\bibnamefont {Hünecke}}, \bibinfo {author}
  {\bibfnamefont {M.}~\bibnamefont {Sierka}}, \bibinfo {author} {\bibfnamefont
  {M.}~\bibnamefont {Neuhaus}}, \bibinfo {author} {\bibfnamefont
  {L.}~\bibnamefont {Hedewig}}, \bibinfo {author} {\bibfnamefont
  {B.}~\bibnamefont {Bergues}}, \bibinfo {author} {\bibfnamefont
  {M.}~\bibnamefont {Alharbi}}, \bibinfo {author} {\bibfnamefont
  {H.}~\bibnamefont {ALQahtani}}, \bibinfo {author} {\bibfnamefont {A.~M.}\
  \bibnamefont {Azzeer}}, \bibinfo {author} {\bibfnamefont {S.}~\bibnamefont
  {Gräfe}}, \bibinfo {author} {\bibfnamefont {M.~F.}\ \bibnamefont {Kling}},
  \bibinfo {author} {\bibfnamefont {A.~F.}\ \bibnamefont {Alharbi}}, \ and\
  \bibinfo {author} {\bibfnamefont {Z.}~\bibnamefont {Wang}},\ }\bibfield
  {title} {\enquote {\bibinfo {title} {Resonance {Effect} in {Brunel}
  {Harmonic} {Generation} in {Thin} {Film} {Organic} {Semiconductors}},}\
  }\href {\doibase 10.1002/adom.202203070} {\bibfield  {journal} {\bibinfo
  {journal} {Advanced Optical Materials}\ }\textbf {\bibinfo {volume} {11}},\
  \bibinfo {pages} {2203070} (\bibinfo {year} {2023})}\BibitemShut {NoStop}%
\bibitem [{\citenamefont {Luu}\ \emph {et~al.}(2018)\citenamefont {Luu},
  \citenamefont {Yin}, \citenamefont {Jain}, \citenamefont {Gaumnitz},
  \citenamefont {Pertot}, \citenamefont {Ma},\ and\ \citenamefont
  {Wörner}}]{luu_extremeultraviolet_2018}%
  \BibitemOpen
  \bibfield  {author} {\bibinfo {author} {\bibfnamefont {T.~T.}\ \bibnamefont
  {Luu}}, \bibinfo {author} {\bibfnamefont {Z.}~\bibnamefont {Yin}}, \bibinfo
  {author} {\bibfnamefont {A.}~\bibnamefont {Jain}}, \bibinfo {author}
  {\bibfnamefont {T.}~\bibnamefont {Gaumnitz}}, \bibinfo {author}
  {\bibfnamefont {Y.}~\bibnamefont {Pertot}}, \bibinfo {author} {\bibfnamefont
  {J.}~\bibnamefont {Ma}}, \ and\ \bibinfo {author} {\bibfnamefont {H.~J.}\
  \bibnamefont {Wörner}},\ }\bibfield  {title} {\enquote {\bibinfo {title}
  {Extreme–ultraviolet high–harmonic generation in liquids},}\ }\href
  {\doibase 10.1038/s41467-018-06040-4} {\bibfield  {journal} {\bibinfo
  {journal} {Nature Communications}\ }\textbf {\bibinfo {volume} {9}},\
  \bibinfo {pages} {3723} (\bibinfo {year} {2018})}\BibitemShut {NoStop}%
\bibitem [{\citenamefont {Nakagawa}\ \emph {et~al.}(2022)\citenamefont
  {Nakagawa}, \citenamefont {Hirori}, \citenamefont {Sato}, \citenamefont
  {Tahara}, \citenamefont {Sekiguchi}, \citenamefont {Yumoto}, \citenamefont
  {Saruyama}, \citenamefont {Sato}, \citenamefont {Teranishi},\ and\
  \citenamefont {Kanemitsu}}]{nakagawa_2022}%
  \BibitemOpen
  \bibfield  {author} {\bibinfo {author} {\bibfnamefont {K.}~\bibnamefont
  {Nakagawa}}, \bibinfo {author} {\bibfnamefont {H.}~\bibnamefont {Hirori}},
  \bibinfo {author} {\bibfnamefont {S.~A.}\ \bibnamefont {Sato}}, \bibinfo
  {author} {\bibfnamefont {H.}~\bibnamefont {Tahara}}, \bibinfo {author}
  {\bibfnamefont {F.}~\bibnamefont {Sekiguchi}}, \bibinfo {author}
  {\bibfnamefont {G.}~\bibnamefont {Yumoto}}, \bibinfo {author} {\bibfnamefont
  {M.}~\bibnamefont {Saruyama}}, \bibinfo {author} {\bibfnamefont
  {R.}~\bibnamefont {Sato}}, \bibinfo {author} {\bibfnamefont {T.}~\bibnamefont
  {Teranishi}}, \ and\ \bibinfo {author} {\bibfnamefont {Y.}~\bibnamefont
  {Kanemitsu}},\ }\bibfield  {title} {\enquote {\bibinfo {title}
  {Size-controlled quantum dots reveal the impact of intraband transitions on
  high-order harmonic generation in solids},}\ }\href {\doibase
  10.1038/s41567-022-01639-3} {\bibfield  {journal} {\bibinfo  {journal}
  {Nature Physics}\ } (\bibinfo {year} {2022}),\
  10.1038/s41567-022-01639-3}\BibitemShut {NoStop}%
\bibitem [{\citenamefont {Gopalakrishna}\ \emph {et~al.}(2023)\citenamefont
  {Gopalakrishna}, \citenamefont {Baruah}, \citenamefont {Hünecke},
  \citenamefont {Korolev}, \citenamefont {Thümmler}, \citenamefont {Croy},
  \citenamefont {Richter}, \citenamefont {Yahyaei}, \citenamefont {Hollinger},
  \citenamefont {Shumakova}, \citenamefont {Uschmann}, \citenamefont
  {Marschner}, \citenamefont {Zürch}, \citenamefont {Reichardt}, \citenamefont
  {Undisz}, \citenamefont {Dellith}, \citenamefont {Pugžlys}, \citenamefont
  {Baltuška}, \citenamefont {Spielmann}, \citenamefont {Peschel},
  \citenamefont {Gräfe}, \citenamefont {Wächtler},\ and\ \citenamefont
  {Kartashov}}]{gopalakrishna_2023}%
  \BibitemOpen
  \bibfield  {author} {\bibinfo {author} {\bibfnamefont {H.~N.}\ \bibnamefont
  {Gopalakrishna}}, \bibinfo {author} {\bibfnamefont {R.}~\bibnamefont
  {Baruah}}, \bibinfo {author} {\bibfnamefont {C.}~\bibnamefont {Hünecke}},
  \bibinfo {author} {\bibfnamefont {V.}~\bibnamefont {Korolev}}, \bibinfo
  {author} {\bibfnamefont {M.}~\bibnamefont {Thümmler}}, \bibinfo {author}
  {\bibfnamefont {A.}~\bibnamefont {Croy}}, \bibinfo {author} {\bibfnamefont
  {M.}~\bibnamefont {Richter}}, \bibinfo {author} {\bibfnamefont
  {F.}~\bibnamefont {Yahyaei}}, \bibinfo {author} {\bibfnamefont
  {R.}~\bibnamefont {Hollinger}}, \bibinfo {author} {\bibfnamefont
  {V.}~\bibnamefont {Shumakova}}, \bibinfo {author} {\bibfnamefont
  {I.}~\bibnamefont {Uschmann}}, \bibinfo {author} {\bibfnamefont
  {H.}~\bibnamefont {Marschner}}, \bibinfo {author} {\bibfnamefont
  {M.}~\bibnamefont {Zürch}}, \bibinfo {author} {\bibfnamefont
  {C.}~\bibnamefont {Reichardt}}, \bibinfo {author} {\bibfnamefont
  {A.}~\bibnamefont {Undisz}}, \bibinfo {author} {\bibfnamefont
  {J.}~\bibnamefont {Dellith}}, \bibinfo {author} {\bibfnamefont
  {A.}~\bibnamefont {Pugžlys}}, \bibinfo {author} {\bibfnamefont
  {A.}~\bibnamefont {Baltuška}}, \bibinfo {author} {\bibfnamefont
  {C.}~\bibnamefont {Spielmann}}, \bibinfo {author} {\bibfnamefont
  {U.}~\bibnamefont {Peschel}}, \bibinfo {author} {\bibfnamefont
  {S.}~\bibnamefont {Gräfe}}, \bibinfo {author} {\bibfnamefont
  {M.}~\bibnamefont {Wächtler}}, \ and\ \bibinfo {author} {\bibfnamefont
  {D.}~\bibnamefont {Kartashov}},\ }\bibfield  {title} {\enquote {\bibinfo
  {title} {Tracing spatial confinement in semiconductor quantum dots by
  high-order harmonic generation},}\ }\href {\doibase
  10.1103/PhysRevResearch.5.013128} {\bibfield  {journal} {\bibinfo  {journal}
  {Physical Review Research}\ }\textbf {\bibinfo {volume} {5}},\ \bibinfo
  {pages} {013128} (\bibinfo {year} {2023})}\BibitemShut {NoStop}%
\bibitem [{\citenamefont {Reiff}\ \emph {et~al.}(2020)\citenamefont {Reiff},
  \citenamefont {Joyce}, \citenamefont {Jaro{\'n}-Becker},\ and\ \citenamefont
  {Becker}}]{reiff2020single}%
  \BibitemOpen
  \bibfield  {author} {\bibinfo {author} {\bibfnamefont {R.}~\bibnamefont
  {Reiff}}, \bibinfo {author} {\bibfnamefont {T.}~\bibnamefont {Joyce}},
  \bibinfo {author} {\bibfnamefont {A.}~\bibnamefont {Jaro{\'n}-Becker}}, \
  and\ \bibinfo {author} {\bibfnamefont {A.}~\bibnamefont {Becker}},\
  }\bibfield  {title} {\enquote {\bibinfo {title} {Single-active electron
  calculations of high-order harmonic generation from valence shells in atoms
  for quantitative comparison with tddft calculations},}\ }\href@noop {}
  {\bibfield  {journal} {\bibinfo  {journal} {Journal of Physics
  Communications}\ }\textbf {\bibinfo {volume} {4}},\ \bibinfo {pages} {065011}
  (\bibinfo {year} {2020})}\BibitemShut {NoStop}%
\bibitem [{\citenamefont {Chu}\ and\ \citenamefont {Chu}(2001)}]{chu_2001}%
  \BibitemOpen
  \bibfield  {author} {\bibinfo {author} {\bibfnamefont {X.}~\bibnamefont
  {Chu}}\ and\ \bibinfo {author} {\bibfnamefont {S.-I.}\ \bibnamefont {Chu}},\
  }\bibfield  {title} {\enquote {\bibinfo {title} {Self-interaction-free
  time-dependent density-functional theory for molecular processes in strong
  fields: High-order harmonic generation of ${\mathrm{h}}_{2}$ in intense laser
  fields},}\ }\href {\doibase 10.1103/PhysRevA.63.023411} {\bibfield  {journal}
  {\bibinfo  {journal} {Phys. Rev. A}\ }\textbf {\bibinfo {volume} {63}},\
  \bibinfo {pages} {023411} (\bibinfo {year} {2001})}\BibitemShut {NoStop}%
\bibitem [{\citenamefont {Tancogne-Dejean}\ and\ \citenamefont
  {Rubio}(2018)}]{tancogne2018atomic}%
  \BibitemOpen
  \bibfield  {author} {\bibinfo {author} {\bibfnamefont {N.}~\bibnamefont
  {Tancogne-Dejean}}\ and\ \bibinfo {author} {\bibfnamefont {A.}~\bibnamefont
  {Rubio}},\ }\bibfield  {title} {\enquote {\bibinfo {title} {Atomic-like
  high-harmonic generation from two-dimensional materials},}\ }\href@noop {}
  {\bibfield  {journal} {\bibinfo  {journal} {Science advances}\ }\textbf
  {\bibinfo {volume} {4}},\ \bibinfo {pages} {eaao5207} (\bibinfo {year}
  {2018})}\BibitemShut {NoStop}%
\bibitem [{\citenamefont {Lindberg}\ and\ \citenamefont
  {Koch}(1988)}]{lindberg_1988}%
  \BibitemOpen
  \bibfield  {author} {\bibinfo {author} {\bibfnamefont {M.}~\bibnamefont
  {Lindberg}}\ and\ \bibinfo {author} {\bibfnamefont {S.~W.}\ \bibnamefont
  {Koch}},\ }\bibfield  {title} {\enquote {\bibinfo {title} {Effective {Bloch}
  equations for semiconductors},}\ }\href {\doibase 10.1103/PhysRevB.38.3342}
  {\bibfield  {journal} {\bibinfo  {journal} {Physical Review B}\ }\textbf
  {\bibinfo {volume} {38}},\ \bibinfo {pages} {3342--3350} (\bibinfo {year}
  {1988})}\BibitemShut {NoStop}%
\bibitem [{\citenamefont {Silva}, \citenamefont {Martín},\ and\ \citenamefont
  {Ivanov}(2019)}]{silva_2019}%
  \BibitemOpen
  \bibfield  {author} {\bibinfo {author} {\bibfnamefont {R.~E.~F.}\
  \bibnamefont {Silva}}, \bibinfo {author} {\bibfnamefont {F.}~\bibnamefont
  {Martín}}, \ and\ \bibinfo {author} {\bibfnamefont {M.}~\bibnamefont
  {Ivanov}},\ }\bibfield  {title} {\enquote {\bibinfo {title} {High harmonic
  generation in crystals using maximally localized {Wannier} functions},}\
  }\href {\doibase 10.1103/PhysRevB.100.195201} {\bibfield  {journal} {\bibinfo
   {journal} {Physical Review B}\ }\textbf {\bibinfo {volume} {100}},\ \bibinfo
  {pages} {195201} (\bibinfo {year} {2019})}\BibitemShut {NoStop}%
\bibitem [{\citenamefont {Jiang}\ \emph {et~al.}(2018)\citenamefont {Jiang},
  \citenamefont {Chen}, \citenamefont {Wei}, \citenamefont {Yu}, \citenamefont
  {Lu},\ and\ \citenamefont {Lin}}]{jiang_2018}%
  \BibitemOpen
  \bibfield  {author} {\bibinfo {author} {\bibfnamefont {S.}~\bibnamefont
  {Jiang}}, \bibinfo {author} {\bibfnamefont {J.}~\bibnamefont {Chen}},
  \bibinfo {author} {\bibfnamefont {H.}~\bibnamefont {Wei}}, \bibinfo {author}
  {\bibfnamefont {C.}~\bibnamefont {Yu}}, \bibinfo {author} {\bibfnamefont
  {R.}~\bibnamefont {Lu}}, \ and\ \bibinfo {author} {\bibfnamefont {C.~D.}\
  \bibnamefont {Lin}},\ }\bibfield  {title} {\enquote {\bibinfo {title} {Role
  of the {Transition} {Dipole} {Amplitude} and {Phase} on the {Generation} of
  {Odd} and {Even} {High}-{Order} {Harmonics} in {Crystals}},}\ }\href
  {\doibase 10.1103/PhysRevLett.120.253201} {\bibfield  {journal} {\bibinfo
  {journal} {Physical Review Letters}\ }\textbf {\bibinfo {volume} {120}},\
  \bibinfo {pages} {253201} (\bibinfo {year} {2018})}\BibitemShut {NoStop}%
\bibitem [{\citenamefont {Tancogne-Dejean}\ \emph {et~al.}(2017)\citenamefont
  {Tancogne-Dejean}, \citenamefont {Mücke}, \citenamefont {Kärtner},\ and\
  \citenamefont {Rubio}}]{tancogne-dejean_2017_1}%
  \BibitemOpen
  \bibfield  {author} {\bibinfo {author} {\bibfnamefont {N.}~\bibnamefont
  {Tancogne-Dejean}}, \bibinfo {author} {\bibfnamefont {O.~D.}\ \bibnamefont
  {Mücke}}, \bibinfo {author} {\bibfnamefont {F.~X.}\ \bibnamefont
  {Kärtner}}, \ and\ \bibinfo {author} {\bibfnamefont {A.}~\bibnamefont
  {Rubio}},\ }\bibfield  {title} {\enquote {\bibinfo {title} {Impact of the
  {Electronic} {Band} {Structure} in {High}-{Harmonic} {Generation} {Spectra}
  of {Solids}},}\ }\href {\doibase 10.1103/PhysRevLett.118.087403} {\bibfield
  {journal} {\bibinfo  {journal} {Physical Review Letters}\ }\textbf {\bibinfo
  {volume} {118}},\ \bibinfo {pages} {087403} (\bibinfo {year}
  {2017})}\BibitemShut {NoStop}%
\bibitem [{\citenamefont {Peschel}\ \emph {et~al.}(2022)\citenamefont
  {Peschel}, \citenamefont {Thümmler}, \citenamefont {Lettau}, \citenamefont
  {Gräfe},\ and\ \citenamefont {Busch}}]{peschel_2022}%
  \BibitemOpen
  \bibfield  {author} {\bibinfo {author} {\bibfnamefont {U.}~\bibnamefont
  {Peschel}}, \bibinfo {author} {\bibfnamefont {M.}~\bibnamefont {Thümmler}},
  \bibinfo {author} {\bibfnamefont {T.}~\bibnamefont {Lettau}}, \bibinfo
  {author} {\bibfnamefont {S.}~\bibnamefont {Gräfe}}, \ and\ \bibinfo {author}
  {\bibfnamefont {K.}~\bibnamefont {Busch}},\ }\bibfield  {title} {\enquote
  {\bibinfo {title} {Two-particle tight-binding description of higher-harmonic
  generation in semiconductor nanostructures},}\ }\href {\doibase
  10.1103/PhysRevB.106.245307} {\bibfield  {journal} {\bibinfo  {journal}
  {Physical Review B}\ }\textbf {\bibinfo {volume} {106}},\ \bibinfo {pages}
  {245307} (\bibinfo {year} {2022})}\BibitemShut {NoStop}%
\bibitem [{\citenamefont {Marzari}\ and\ \citenamefont
  {Vanderbilt}(1997)}]{marzari_1997}%
  \BibitemOpen
  \bibfield  {author} {\bibinfo {author} {\bibfnamefont {N.}~\bibnamefont
  {Marzari}}\ and\ \bibinfo {author} {\bibfnamefont {D.}~\bibnamefont
  {Vanderbilt}},\ }\bibfield  {title} {\enquote {\bibinfo {title} {Maximally
  localized generalized {Wannier} functions for composite energy bands},}\
  }\href {\doibase 10.1103/PhysRevB.56.12847} {\bibfield  {journal} {\bibinfo
  {journal} {Physical Review B}\ }\textbf {\bibinfo {volume} {56}},\ \bibinfo
  {pages} {12847--12865} (\bibinfo {year} {1997})}\BibitemShut {NoStop}%
\bibitem [{\citenamefont {Marzari}\ \emph {et~al.}(2012)\citenamefont
  {Marzari}, \citenamefont {Mostofi}, \citenamefont {Yates}, \citenamefont
  {Souza},\ and\ \citenamefont {Vanderbilt}}]{marzari_2012}%
  \BibitemOpen
  \bibfield  {author} {\bibinfo {author} {\bibfnamefont {N.}~\bibnamefont
  {Marzari}}, \bibinfo {author} {\bibfnamefont {A.~A.}\ \bibnamefont
  {Mostofi}}, \bibinfo {author} {\bibfnamefont {J.~R.}\ \bibnamefont {Yates}},
  \bibinfo {author} {\bibfnamefont {I.}~\bibnamefont {Souza}}, \ and\ \bibinfo
  {author} {\bibfnamefont {D.}~\bibnamefont {Vanderbilt}},\ }\bibfield  {title}
  {\enquote {\bibinfo {title} {Maximally localized {Wannier} functions:
  {Theory} and applications},}\ }\href {\doibase 10.1103/RevModPhys.84.1419}
  {\bibfield  {journal} {\bibinfo  {journal} {Reviews of Modern Physics}\
  }\textbf {\bibinfo {volume} {84}},\ \bibinfo {pages} {1419--1475} (\bibinfo
  {year} {2012})}\BibitemShut {NoStop}%
\bibitem [{\citenamefont {Thümmler}\ \emph {et~al.}(2025)\citenamefont
  {Thümmler}, \citenamefont {Lettau}, \citenamefont {Croy}, \citenamefont
  {Peschel},\ and\ \citenamefont {Gräfe}}]{thummler_semiconductor_2025}%
  \BibitemOpen
  \bibfield  {author} {\bibinfo {author} {\bibfnamefont {M.}~\bibnamefont
  {Thümmler}}, \bibinfo {author} {\bibfnamefont {T.}~\bibnamefont {Lettau}},
  \bibinfo {author} {\bibfnamefont {A.}~\bibnamefont {Croy}}, \bibinfo {author}
  {\bibfnamefont {U.}~\bibnamefont {Peschel}}, \ and\ \bibinfo {author}
  {\bibfnamefont {S.}~\bibnamefont {Gräfe}},\ }\bibfield  {title} {\enquote
  {\bibinfo {title} {Semiconductor {Bloch} equations in {Wannier} gauge with
  well-behaved dephasing},}\ }\href {\doibase 10.1016/j.cpc.2025.109958}
  {\bibfield  {journal} {\bibinfo  {journal} {Computer Physics Communications}\
  ,\ \bibinfo {pages} {109958}} (\bibinfo {year} {2025})}\BibitemShut {NoStop}%
\bibitem [{\citenamefont {Yue}\ and\ \citenamefont {Gaarde}(2022)}]{yue_2022}%
  \BibitemOpen
  \bibfield  {author} {\bibinfo {author} {\bibfnamefont {L.}~\bibnamefont
  {Yue}}\ and\ \bibinfo {author} {\bibfnamefont {M.~B.}\ \bibnamefont
  {Gaarde}},\ }\bibfield  {title} {\enquote {\bibinfo {title} {Introduction to
  theory of high-harmonic generation in solids: tutorial},}\ }\href {\doibase
  10.1364/JOSAB.448602} {\bibfield  {journal} {\bibinfo  {journal} {Journal of
  the Optical Society of America B}\ }\textbf {\bibinfo {volume} {39}},\
  \bibinfo {pages} {535} (\bibinfo {year} {2022})}\BibitemShut {NoStop}%
\bibitem [{\citenamefont {Liu}\ \emph {et~al.}(2016)\citenamefont {Liu},
  \citenamefont {Zheng}, \citenamefont {Zeng},\ and\ \citenamefont
  {Li}}]{liu_2016}%
  \BibitemOpen
  \bibfield  {author} {\bibinfo {author} {\bibfnamefont {C.}~\bibnamefont
  {Liu}}, \bibinfo {author} {\bibfnamefont {Y.}~\bibnamefont {Zheng}}, \bibinfo
  {author} {\bibfnamefont {Z.}~\bibnamefont {Zeng}}, \ and\ \bibinfo {author}
  {\bibfnamefont {R.}~\bibnamefont {Li}},\ }\bibfield  {title} {\enquote
  {\bibinfo {title} {Effect of elliptical polarization of driving field on
  high-order-harmonic generation in semiconductor zno},}\ }\href {\doibase
  10.1103/PhysRevA.93.043806} {\bibfield  {journal} {\bibinfo  {journal} {Phys.
  Rev. A}\ }\textbf {\bibinfo {volume} {93}},\ \bibinfo {pages} {043806}
  (\bibinfo {year} {2016})}\BibitemShut {NoStop}%
\bibitem [{\citenamefont {Hollinger}\ \emph {et~al.}(2021)\citenamefont
  {Hollinger}, \citenamefont {Herrmann}, \citenamefont {Korolev}, \citenamefont
  {Zapf}, \citenamefont {Shumakova}, \citenamefont {Röder}, \citenamefont
  {Uschmann}, \citenamefont {Pugžlys}, \citenamefont {Baltuška},
  \citenamefont {Zürch}, \citenamefont {Ronning}, \citenamefont {Spielmann},\
  and\ \citenamefont {Kartashov}}]{nano11010004}%
  \BibitemOpen
  \bibfield  {author} {\bibinfo {author} {\bibfnamefont {R.}~\bibnamefont
  {Hollinger}}, \bibinfo {author} {\bibfnamefont {P.}~\bibnamefont {Herrmann}},
  \bibinfo {author} {\bibfnamefont {V.}~\bibnamefont {Korolev}}, \bibinfo
  {author} {\bibfnamefont {M.}~\bibnamefont {Zapf}}, \bibinfo {author}
  {\bibfnamefont {V.}~\bibnamefont {Shumakova}}, \bibinfo {author}
  {\bibfnamefont {R.}~\bibnamefont {Röder}}, \bibinfo {author} {\bibfnamefont
  {I.}~\bibnamefont {Uschmann}}, \bibinfo {author} {\bibfnamefont
  {A.}~\bibnamefont {Pugžlys}}, \bibinfo {author} {\bibfnamefont
  {A.}~\bibnamefont {Baltuška}}, \bibinfo {author} {\bibfnamefont
  {M.}~\bibnamefont {Zürch}}, \bibinfo {author} {\bibfnamefont
  {C.}~\bibnamefont {Ronning}}, \bibinfo {author} {\bibfnamefont
  {C.}~\bibnamefont {Spielmann}}, \ and\ \bibinfo {author} {\bibfnamefont
  {D.}~\bibnamefont {Kartashov}},\ }\bibfield  {title} {\enquote {\bibinfo
  {title} {Polarization dependent excitation and high harmonic generation from
  intense mid-ir laser pulses in zno},}\ }\href {\doibase 10.3390/nano11010004}
  {\bibfield  {journal} {\bibinfo  {journal} {Nanomaterials}\ }\textbf
  {\bibinfo {volume} {11}} (\bibinfo {year} {2021}),\
  10.3390/nano11010004}\BibitemShut {NoStop}%
\bibitem [{\citenamefont {Pizzi}\ \emph {et~al.}(2020)\citenamefont {Pizzi},
  \citenamefont {Vitale}, \citenamefont {Arita}, \citenamefont {Blügel},
  \citenamefont {Freimuth}, \citenamefont {Géranton}, \citenamefont
  {Gibertini}, \citenamefont {Gresch}, \citenamefont {Johnson}, \citenamefont
  {Koretsune}, \citenamefont {Ibañez-Azpiroz}, \citenamefont {Lee},
  \citenamefont {Lihm}, \citenamefont {Marchand}, \citenamefont {Marrazzo},
  \citenamefont {Mokrousov}, \citenamefont {Mustafa}, \citenamefont {Nohara},
  \citenamefont {Nomura}, \citenamefont {Paulatto}, \citenamefont {Poncé},
  \citenamefont {Ponweiser}, \citenamefont {Qiao}, \citenamefont {Thöle},
  \citenamefont {Tsirkin}, \citenamefont {Wierzbowska}, \citenamefont
  {Marzari}, \citenamefont {Vanderbilt}, \citenamefont {Souza}, \citenamefont
  {Mostofi},\ and\ \citenamefont {Yates}}]{pizzi_2020}%
  \BibitemOpen
  \bibfield  {author} {\bibinfo {author} {\bibfnamefont {G.}~\bibnamefont
  {Pizzi}}, \bibinfo {author} {\bibfnamefont {V.}~\bibnamefont {Vitale}},
  \bibinfo {author} {\bibfnamefont {R.}~\bibnamefont {Arita}}, \bibinfo
  {author} {\bibfnamefont {S.}~\bibnamefont {Blügel}}, \bibinfo {author}
  {\bibfnamefont {F.}~\bibnamefont {Freimuth}}, \bibinfo {author}
  {\bibfnamefont {G.}~\bibnamefont {Géranton}}, \bibinfo {author}
  {\bibfnamefont {M.}~\bibnamefont {Gibertini}}, \bibinfo {author}
  {\bibfnamefont {D.}~\bibnamefont {Gresch}}, \bibinfo {author} {\bibfnamefont
  {C.}~\bibnamefont {Johnson}}, \bibinfo {author} {\bibfnamefont
  {T.}~\bibnamefont {Koretsune}}, \bibinfo {author} {\bibfnamefont
  {J.}~\bibnamefont {Ibañez-Azpiroz}}, \bibinfo {author} {\bibfnamefont
  {H.}~\bibnamefont {Lee}}, \bibinfo {author} {\bibfnamefont {J.-M.}\
  \bibnamefont {Lihm}}, \bibinfo {author} {\bibfnamefont {D.}~\bibnamefont
  {Marchand}}, \bibinfo {author} {\bibfnamefont {A.}~\bibnamefont {Marrazzo}},
  \bibinfo {author} {\bibfnamefont {Y.}~\bibnamefont {Mokrousov}}, \bibinfo
  {author} {\bibfnamefont {J.~I.}\ \bibnamefont {Mustafa}}, \bibinfo {author}
  {\bibfnamefont {Y.}~\bibnamefont {Nohara}}, \bibinfo {author} {\bibfnamefont
  {Y.}~\bibnamefont {Nomura}}, \bibinfo {author} {\bibfnamefont
  {L.}~\bibnamefont {Paulatto}}, \bibinfo {author} {\bibfnamefont
  {S.}~\bibnamefont {Poncé}}, \bibinfo {author} {\bibfnamefont
  {T.}~\bibnamefont {Ponweiser}}, \bibinfo {author} {\bibfnamefont
  {J.}~\bibnamefont {Qiao}}, \bibinfo {author} {\bibfnamefont {F.}~\bibnamefont
  {Thöle}}, \bibinfo {author} {\bibfnamefont {S.~S.}\ \bibnamefont {Tsirkin}},
  \bibinfo {author} {\bibfnamefont {M.}~\bibnamefont {Wierzbowska}}, \bibinfo
  {author} {\bibfnamefont {N.}~\bibnamefont {Marzari}}, \bibinfo {author}
  {\bibfnamefont {D.}~\bibnamefont {Vanderbilt}}, \bibinfo {author}
  {\bibfnamefont {I.}~\bibnamefont {Souza}}, \bibinfo {author} {\bibfnamefont
  {A.~A.}\ \bibnamefont {Mostofi}}, \ and\ \bibinfo {author} {\bibfnamefont
  {J.~R.}\ \bibnamefont {Yates}},\ }\bibfield  {title} {\enquote {\bibinfo
  {title} {Wannier90 as a community code: new features and applications},}\
  }\href {\doibase 10.1088/1361-648X/ab51ff} {\bibfield  {journal} {\bibinfo
  {journal} {Journal of Physics: Condensed Matter}\ }\textbf {\bibinfo {volume}
  {32}},\ \bibinfo {pages} {165902} (\bibinfo {year} {2020})}\BibitemShut
  {NoStop}%
\bibitem [{\citenamefont {Souza}, \citenamefont {Marzari},\ and\ \citenamefont
  {Vanderbilt}(2001)}]{souza_2001}%
  \BibitemOpen
  \bibfield  {author} {\bibinfo {author} {\bibfnamefont {I.}~\bibnamefont
  {Souza}}, \bibinfo {author} {\bibfnamefont {N.}~\bibnamefont {Marzari}}, \
  and\ \bibinfo {author} {\bibfnamefont {D.}~\bibnamefont {Vanderbilt}},\
  }\bibfield  {title} {\enquote {\bibinfo {title} {Maximally localized
  {Wannier} functions for entangled energy bands},}\ }\href {\doibase
  10.1103/PhysRevB.65.035109} {\bibfield  {journal} {\bibinfo  {journal}
  {Physical Review B}\ }\textbf {\bibinfo {volume} {65}},\ \bibinfo {pages}
  {035109} (\bibinfo {year} {2001})}\BibitemShut {NoStop}%
\bibitem [{\citenamefont {Wang}\ \emph {et~al.}(2006)\citenamefont {Wang},
  \citenamefont {Yates}, \citenamefont {Souza},\ and\ \citenamefont
  {Vanderbilt}}]{wang_ab_2006}%
  \BibitemOpen
  \bibfield  {author} {\bibinfo {author} {\bibfnamefont {X.}~\bibnamefont
  {Wang}}, \bibinfo {author} {\bibfnamefont {J.~R.}\ \bibnamefont {Yates}},
  \bibinfo {author} {\bibfnamefont {I.}~\bibnamefont {Souza}}, \ and\ \bibinfo
  {author} {\bibfnamefont {D.}~\bibnamefont {Vanderbilt}},\ }\bibfield  {title}
  {\enquote {\bibinfo {title} {\textit{{Ab} initio} calculation of the
  anomalous {Hall} conductivity by {Wannier} interpolation},}\ }\href {\doibase
  10.1103/PhysRevB.74.195118} {\bibfield  {journal} {\bibinfo  {journal}
  {Physical Review B}\ }\textbf {\bibinfo {volume} {74}},\ \bibinfo {pages}
  {195118} (\bibinfo {year} {2006})}\BibitemShut {NoStop}%
\bibitem [{\citenamefont {Meulenberg}\ \emph {et~al.}(2009)\citenamefont
  {Meulenberg}, \citenamefont {Lee}, \citenamefont {Wolcott}, \citenamefont
  {Zhang}, \citenamefont {Terminello},\ and\ \citenamefont {van
  Buuren}}]{meulenberg_determination_2009}%
  \BibitemOpen
  \bibfield  {author} {\bibinfo {author} {\bibfnamefont {R.~W.}\ \bibnamefont
  {Meulenberg}}, \bibinfo {author} {\bibfnamefont {J.~R.}\ \bibnamefont {Lee}},
  \bibinfo {author} {\bibfnamefont {A.}~\bibnamefont {Wolcott}}, \bibinfo
  {author} {\bibfnamefont {J.~Z.}\ \bibnamefont {Zhang}}, \bibinfo {author}
  {\bibfnamefont {L.~J.}\ \bibnamefont {Terminello}}, \ and\ \bibinfo {author}
  {\bibfnamefont {T.}~\bibnamefont {van Buuren}},\ }\bibfield  {title}
  {\enquote {\bibinfo {title} {Determination of the {Exciton} {Binding}
  {Energy} in {CdSe} {Quantum} {Dots}},}\ }\href {\doibase 10.1021/nn8006916}
  {\bibfield  {journal} {\bibinfo  {journal} {ACS Nano}\ }\textbf {\bibinfo
  {volume} {3}},\ \bibinfo {pages} {325--330} (\bibinfo {year} {2009})},\
  \bibinfo {note} {publisher: American Chemical Society (ACS)}\BibitemShut
  {NoStop}%
\bibitem [{\citenamefont {Corkum}(1993)}]{corkum_plasma_1993}%
  \BibitemOpen
  \bibfield  {author} {\bibinfo {author} {\bibfnamefont {P.~B.}\ \bibnamefont
  {Corkum}},\ }\bibfield  {title} {\enquote {\bibinfo {title} {Plasma
  perspective on strong field multiphoton ionization},}\ }\href {\doibase
  10.1103/PhysRevLett.71.1994} {\bibfield  {journal} {\bibinfo  {journal}
  {Physical Review Letters}\ }\textbf {\bibinfo {volume} {71}},\ \bibinfo
  {pages} {1994--1997} (\bibinfo {year} {1993})}\BibitemShut {NoStop}%
\bibitem [{\citenamefont {Tao}\ and\ \citenamefont
  {Scrinzi}(2012)}]{tao_photo-electron_2012}%
  \BibitemOpen
  \bibfield  {author} {\bibinfo {author} {\bibfnamefont {L.}~\bibnamefont
  {Tao}}\ and\ \bibinfo {author} {\bibfnamefont {A.}~\bibnamefont {Scrinzi}},\
  }\bibfield  {title} {\enquote {\bibinfo {title} {Photo-electron momentum
  spectra from minimal volumes: the time-dependent surface flux method},}\
  }\href {\doibase 10.1088/1367-2630/14/1/013021} {\bibfield  {journal}
  {\bibinfo  {journal} {New Journal of Physics}\ }\textbf {\bibinfo {volume}
  {14}},\ \bibinfo {pages} {013021} (\bibinfo {year} {2012})}\BibitemShut
  {NoStop}%
\bibitem [{\citenamefont {Harrison}(2012)}]{harrison2012electronic}%
  \BibitemOpen
  \bibfield  {author} {\bibinfo {author} {\bibfnamefont {W.~A.}\ \bibnamefont
  {Harrison}},\ }\href@noop {} {\emph {\bibinfo {title} {Electronic structure
  and the properties of solids: the physics of the chemical bond}}}\ (\bibinfo
  {publisher} {Courier Corporation},\ \bibinfo {year} {2012})\BibitemShut
  {NoStop}%
\bibitem [{\citenamefont {Ahnert}\ and\ \citenamefont
  {Mulansky}(2011)}]{ahnert_2011}%
  \BibitemOpen
  \bibfield  {author} {\bibinfo {author} {\bibfnamefont {K.}~\bibnamefont
  {Ahnert}}\ and\ \bibinfo {author} {\bibfnamefont {M.}~\bibnamefont
  {Mulansky}},\ }\bibfield  {title} {\enquote {\bibinfo {title} {Odeint -
  {Solving} ordinary differential equations in {C++}},}\ }\href {\doibase
  10.48550/ARXIV.1110.3397} {\bibfield  {journal} {\bibinfo  {journal} {arxiv}\
  } (\bibinfo {year} {2011}),\ 10.48550/ARXIV.1110.3397}\BibitemShut {NoStop}%
\bibitem [{\citenamefont {Demidov}\ \emph {et~al.}(2013)\citenamefont
  {Demidov}, \citenamefont {Ahnert}, \citenamefont {Rupp},\ and\ \citenamefont
  {Gottschling}}]{demidov_2013}%
  \BibitemOpen
  \bibfield  {author} {\bibinfo {author} {\bibfnamefont {D.}~\bibnamefont
  {Demidov}}, \bibinfo {author} {\bibfnamefont {K.}~\bibnamefont {Ahnert}},
  \bibinfo {author} {\bibfnamefont {K.}~\bibnamefont {Rupp}}, \ and\ \bibinfo
  {author} {\bibfnamefont {P.}~\bibnamefont {Gottschling}},\ }\bibfield
  {title} {\enquote {\bibinfo {title} {Programming {CUDA} and {OpenCL}: {A}
  {Case} {Study} {Using} {Modern} {C}++ {Libraries}},}\ }\href {\doibase
  10.1137/120903683} {\bibfield  {journal} {\bibinfo  {journal} {SIAM Journal
  on Scientific Computing}\ }\textbf {\bibinfo {volume} {35}},\ \bibinfo
  {pages} {C453--C472} (\bibinfo {year} {2013})}\BibitemShut {NoStop}%
\bibitem [{thu(2024{\natexlab{a}})}]{thuemmler_2024_1}%
  \BibitemOpen
  \href@noop {} {\enquote {\bibinfo {title} {Code of finite-size model},}\
  }\bibinfo {howpublished} {\url{https://git.uni-jena.de/pa74roz/TB-QD-HHG}}
  (\bibinfo {year} {2024}{\natexlab{a}}),\ \bibinfo {note} {version
  1.1}\BibitemShut {NoStop}%
\bibitem [{\citenamefont {Wilhelm}\ \emph {et~al.}(2021)\citenamefont
  {Wilhelm}, \citenamefont {Grössing}, \citenamefont {Seith}, \citenamefont
  {Crewse}, \citenamefont {Nitsch}, \citenamefont {Weigl}, \citenamefont
  {Schmid},\ and\ \citenamefont {Evers}}]{wilhelm_semiconductor_2021}%
  \BibitemOpen
  \bibfield  {author} {\bibinfo {author} {\bibfnamefont {J.}~\bibnamefont
  {Wilhelm}}, \bibinfo {author} {\bibfnamefont {P.}~\bibnamefont {Grössing}},
  \bibinfo {author} {\bibfnamefont {A.}~\bibnamefont {Seith}}, \bibinfo
  {author} {\bibfnamefont {J.}~\bibnamefont {Crewse}}, \bibinfo {author}
  {\bibfnamefont {M.}~\bibnamefont {Nitsch}}, \bibinfo {author} {\bibfnamefont
  {L.}~\bibnamefont {Weigl}}, \bibinfo {author} {\bibfnamefont
  {C.}~\bibnamefont {Schmid}}, \ and\ \bibinfo {author} {\bibfnamefont
  {F.}~\bibnamefont {Evers}},\ }\bibfield  {title} {\enquote {\bibinfo {title}
  {Semiconductor {Bloch}-equations formalism: {Derivation} and application to
  high-harmonic generation from {Dirac} fermions},}\ }\href {\doibase
  10.1103/PhysRevB.103.125419} {\bibfield  {journal} {\bibinfo  {journal}
  {Physical Review B}\ }\textbf {\bibinfo {volume} {103}},\ \bibinfo {pages}
  {125419} (\bibinfo {year} {2021})}\BibitemShut {NoStop}%
\bibitem [{\citenamefont {Larmor}(1897)}]{larmor_lxiii_1897}%
  \BibitemOpen
  \bibfield  {author} {\bibinfo {author} {\bibfnamefont {J.}~\bibnamefont
  {Larmor}},\ }\bibfield  {title} {\enquote {\bibinfo {title} {{LXIII}.
  \textit{{On} the theory of the magnetic influence on spectra; and on the
  radiation from moving ions}},}\ }\href {\doibase 10.1080/14786449708621095}
  {\bibfield  {journal} {\bibinfo  {journal} {The London, Edinburgh, and Dublin
  Philosophical Magazine and Journal of Science}\ }\textbf {\bibinfo {volume}
  {44}},\ \bibinfo {pages} {503--512} (\bibinfo {year} {1897})},\ \bibinfo
  {note} {publisher: Informa UK Limited}\BibitemShut {NoStop}%
\bibitem [{thu(2024{\natexlab{b}})}]{thuemmler_2024_2}%
  \BibitemOpen
  \href@noop {} {\enquote {\bibinfo {title} {Code of real-space periodic
  model},}\ }\bibinfo {howpublished}
  {\url{https://git.uni-jena.de/pa74roz/periodic-wannier-tb}} (\bibinfo {year}
  {2024}{\natexlab{b}}),\ \bibinfo {note} {version 1.0}\BibitemShut {NoStop}%
\bibitem [{\citenamefont {Greenwood}(1958)}]{greenwood_1958}%
  \BibitemOpen
  \bibfield  {author} {\bibinfo {author} {\bibfnamefont {D.~A.}\ \bibnamefont
  {Greenwood}},\ }\bibfield  {title} {\enquote {\bibinfo {title} {The
  {Boltzmann} {Equation} in the {Theory} of {Electrical} {Conduction} in
  {Metals}},}\ }\href {\doibase 10.1088/0370-1328/71/4/306} {\bibfield
  {journal} {\bibinfo  {journal} {Proceedings of the Physical Society}\
  }\textbf {\bibinfo {volume} {71}},\ \bibinfo {pages} {585--596} (\bibinfo
  {year} {1958})}\BibitemShut {NoStop}%
\bibitem [{\citenamefont {Giannozzi}\ \emph {et~al.}(2009)\citenamefont
  {Giannozzi}, \citenamefont {Baroni}, \citenamefont {Bonini}, \citenamefont
  {Calandra}, \citenamefont {Car}, \citenamefont {Cavazzoni}, \citenamefont
  {Ceresoli}, \citenamefont {Chiarotti}, \citenamefont {Cococcioni},
  \citenamefont {Dabo}, \citenamefont {Dal~Corso}, \citenamefont
  {de~Gironcoli}, \citenamefont {Fabris}, \citenamefont {Fratesi},
  \citenamefont {Gebauer}, \citenamefont {Gerstmann}, \citenamefont
  {Gougoussis}, \citenamefont {Kokalj}, \citenamefont {Lazzeri}, \citenamefont
  {Martin-Samos}, \citenamefont {Marzari}, \citenamefont {Mauri}, \citenamefont
  {Mazzarello}, \citenamefont {Paolini}, \citenamefont {Pasquarello},
  \citenamefont {Paulatto}, \citenamefont {Sbraccia}, \citenamefont {Scandolo},
  \citenamefont {Sclauzero}, \citenamefont {Seitsonen}, \citenamefont
  {Smogunov}, \citenamefont {Umari},\ and\ \citenamefont
  {Wentzcovitch}}]{giannozzi_2009}%
  \BibitemOpen
  \bibfield  {author} {\bibinfo {author} {\bibfnamefont {P.}~\bibnamefont
  {Giannozzi}}, \bibinfo {author} {\bibfnamefont {S.}~\bibnamefont {Baroni}},
  \bibinfo {author} {\bibfnamefont {N.}~\bibnamefont {Bonini}}, \bibinfo
  {author} {\bibfnamefont {M.}~\bibnamefont {Calandra}}, \bibinfo {author}
  {\bibfnamefont {R.}~\bibnamefont {Car}}, \bibinfo {author} {\bibfnamefont
  {C.}~\bibnamefont {Cavazzoni}}, \bibinfo {author} {\bibfnamefont
  {D.}~\bibnamefont {Ceresoli}}, \bibinfo {author} {\bibfnamefont {G.~L.}\
  \bibnamefont {Chiarotti}}, \bibinfo {author} {\bibfnamefont {M.}~\bibnamefont
  {Cococcioni}}, \bibinfo {author} {\bibfnamefont {I.}~\bibnamefont {Dabo}},
  \bibinfo {author} {\bibfnamefont {A.}~\bibnamefont {Dal~Corso}}, \bibinfo
  {author} {\bibfnamefont {S.}~\bibnamefont {de~Gironcoli}}, \bibinfo {author}
  {\bibfnamefont {S.}~\bibnamefont {Fabris}}, \bibinfo {author} {\bibfnamefont
  {G.}~\bibnamefont {Fratesi}}, \bibinfo {author} {\bibfnamefont
  {R.}~\bibnamefont {Gebauer}}, \bibinfo {author} {\bibfnamefont
  {U.}~\bibnamefont {Gerstmann}}, \bibinfo {author} {\bibfnamefont
  {C.}~\bibnamefont {Gougoussis}}, \bibinfo {author} {\bibfnamefont
  {A.}~\bibnamefont {Kokalj}}, \bibinfo {author} {\bibfnamefont
  {M.}~\bibnamefont {Lazzeri}}, \bibinfo {author} {\bibfnamefont
  {L.}~\bibnamefont {Martin-Samos}}, \bibinfo {author} {\bibfnamefont
  {N.}~\bibnamefont {Marzari}}, \bibinfo {author} {\bibfnamefont
  {F.}~\bibnamefont {Mauri}}, \bibinfo {author} {\bibfnamefont
  {R.}~\bibnamefont {Mazzarello}}, \bibinfo {author} {\bibfnamefont
  {S.}~\bibnamefont {Paolini}}, \bibinfo {author} {\bibfnamefont
  {A.}~\bibnamefont {Pasquarello}}, \bibinfo {author} {\bibfnamefont
  {L.}~\bibnamefont {Paulatto}}, \bibinfo {author} {\bibfnamefont
  {C.}~\bibnamefont {Sbraccia}}, \bibinfo {author} {\bibfnamefont
  {S.}~\bibnamefont {Scandolo}}, \bibinfo {author} {\bibfnamefont
  {G.}~\bibnamefont {Sclauzero}}, \bibinfo {author} {\bibfnamefont {A.~P.}\
  \bibnamefont {Seitsonen}}, \bibinfo {author} {\bibfnamefont {A.}~\bibnamefont
  {Smogunov}}, \bibinfo {author} {\bibfnamefont {P.}~\bibnamefont {Umari}}, \
  and\ \bibinfo {author} {\bibfnamefont {R.~M.}\ \bibnamefont {Wentzcovitch}},\
  }\bibfield  {title} {\enquote {\bibinfo {title} {{QUANTUM} {ESPRESSO}: a
  modular and open-source software project for quantum simulations of
  materials},}\ }\href {\doibase 10.1088/0953-8984/21/39/395502} {\bibfield
  {journal} {\bibinfo  {journal} {Journal of Physics: Condensed Matter}\
  }\textbf {\bibinfo {volume} {21}},\ \bibinfo {pages} {395502} (\bibinfo
  {year} {2009})}\BibitemShut {NoStop}%
\bibitem [{\citenamefont {Giannozzi}\ \emph {et~al.}(2017)\citenamefont
  {Giannozzi}, \citenamefont {Andreussi}, \citenamefont {Brumme}, \citenamefont
  {Bunau}, \citenamefont {Buongiorno~Nardelli}, \citenamefont {Calandra},
  \citenamefont {Car}, \citenamefont {Cavazzoni}, \citenamefont {Ceresoli},
  \citenamefont {Cococcioni}, \citenamefont {Colonna}, \citenamefont
  {Carnimeo}, \citenamefont {Dal~Corso}, \citenamefont {de~Gironcoli},
  \citenamefont {Delugas}, \citenamefont {DiStasio}, \citenamefont {Ferretti},
  \citenamefont {Floris}, \citenamefont {Fratesi}, \citenamefont {Fugallo},
  \citenamefont {Gebauer}, \citenamefont {Gerstmann}, \citenamefont {Giustino},
  \citenamefont {Gorni}, \citenamefont {Jia}, \citenamefont {Kawamura},
  \citenamefont {Ko}, \citenamefont {Kokalj}, \citenamefont {Küçükbenli},
  \citenamefont {Lazzeri}, \citenamefont {Marsili}, \citenamefont {Marzari},
  \citenamefont {Mauri}, \citenamefont {Nguyen}, \citenamefont {Nguyen},
  \citenamefont {Otero-de-la Roza}, \citenamefont {Paulatto}, \citenamefont
  {Poncé}, \citenamefont {Rocca}, \citenamefont {Sabatini}, \citenamefont
  {Santra}, \citenamefont {Schlipf}, \citenamefont {Seitsonen}, \citenamefont
  {Smogunov}, \citenamefont {Timrov}, \citenamefont {Thonhauser}, \citenamefont
  {Umari}, \citenamefont {Vast}, \citenamefont {Wu},\ and\ \citenamefont
  {Baroni}}]{giannozzi_2017}%
  \BibitemOpen
  \bibfield  {author} {\bibinfo {author} {\bibfnamefont {P.}~\bibnamefont
  {Giannozzi}}, \bibinfo {author} {\bibfnamefont {O.}~\bibnamefont
  {Andreussi}}, \bibinfo {author} {\bibfnamefont {T.}~\bibnamefont {Brumme}},
  \bibinfo {author} {\bibfnamefont {O.}~\bibnamefont {Bunau}}, \bibinfo
  {author} {\bibfnamefont {M.}~\bibnamefont {Buongiorno~Nardelli}}, \bibinfo
  {author} {\bibfnamefont {M.}~\bibnamefont {Calandra}}, \bibinfo {author}
  {\bibfnamefont {R.}~\bibnamefont {Car}}, \bibinfo {author} {\bibfnamefont
  {C.}~\bibnamefont {Cavazzoni}}, \bibinfo {author} {\bibfnamefont
  {D.}~\bibnamefont {Ceresoli}}, \bibinfo {author} {\bibfnamefont
  {M.}~\bibnamefont {Cococcioni}}, \bibinfo {author} {\bibfnamefont
  {N.}~\bibnamefont {Colonna}}, \bibinfo {author} {\bibfnamefont
  {I.}~\bibnamefont {Carnimeo}}, \bibinfo {author} {\bibfnamefont
  {A.}~\bibnamefont {Dal~Corso}}, \bibinfo {author} {\bibfnamefont
  {S.}~\bibnamefont {de~Gironcoli}}, \bibinfo {author} {\bibfnamefont
  {P.}~\bibnamefont {Delugas}}, \bibinfo {author} {\bibfnamefont {R.~A.}\
  \bibnamefont {DiStasio}}, \bibinfo {author} {\bibfnamefont {A.}~\bibnamefont
  {Ferretti}}, \bibinfo {author} {\bibfnamefont {A.}~\bibnamefont {Floris}},
  \bibinfo {author} {\bibfnamefont {G.}~\bibnamefont {Fratesi}}, \bibinfo
  {author} {\bibfnamefont {G.}~\bibnamefont {Fugallo}}, \bibinfo {author}
  {\bibfnamefont {R.}~\bibnamefont {Gebauer}}, \bibinfo {author} {\bibfnamefont
  {U.}~\bibnamefont {Gerstmann}}, \bibinfo {author} {\bibfnamefont
  {F.}~\bibnamefont {Giustino}}, \bibinfo {author} {\bibfnamefont
  {T.}~\bibnamefont {Gorni}}, \bibinfo {author} {\bibfnamefont
  {J.}~\bibnamefont {Jia}}, \bibinfo {author} {\bibfnamefont {M.}~\bibnamefont
  {Kawamura}}, \bibinfo {author} {\bibfnamefont {H.-Y.}\ \bibnamefont {Ko}},
  \bibinfo {author} {\bibfnamefont {A.}~\bibnamefont {Kokalj}}, \bibinfo
  {author} {\bibfnamefont {E.}~\bibnamefont {Küçükbenli}}, \bibinfo {author}
  {\bibfnamefont {M.}~\bibnamefont {Lazzeri}}, \bibinfo {author} {\bibfnamefont
  {M.}~\bibnamefont {Marsili}}, \bibinfo {author} {\bibfnamefont
  {N.}~\bibnamefont {Marzari}}, \bibinfo {author} {\bibfnamefont
  {F.}~\bibnamefont {Mauri}}, \bibinfo {author} {\bibfnamefont {N.~L.}\
  \bibnamefont {Nguyen}}, \bibinfo {author} {\bibfnamefont {H.-V.}\
  \bibnamefont {Nguyen}}, \bibinfo {author} {\bibfnamefont {A.}~\bibnamefont
  {Otero-de-la Roza}}, \bibinfo {author} {\bibfnamefont {L.}~\bibnamefont
  {Paulatto}}, \bibinfo {author} {\bibfnamefont {S.}~\bibnamefont {Poncé}},
  \bibinfo {author} {\bibfnamefont {D.}~\bibnamefont {Rocca}}, \bibinfo
  {author} {\bibfnamefont {R.}~\bibnamefont {Sabatini}}, \bibinfo {author}
  {\bibfnamefont {B.}~\bibnamefont {Santra}}, \bibinfo {author} {\bibfnamefont
  {M.}~\bibnamefont {Schlipf}}, \bibinfo {author} {\bibfnamefont {A.~P.}\
  \bibnamefont {Seitsonen}}, \bibinfo {author} {\bibfnamefont {A.}~\bibnamefont
  {Smogunov}}, \bibinfo {author} {\bibfnamefont {I.}~\bibnamefont {Timrov}},
  \bibinfo {author} {\bibfnamefont {T.}~\bibnamefont {Thonhauser}}, \bibinfo
  {author} {\bibfnamefont {P.}~\bibnamefont {Umari}}, \bibinfo {author}
  {\bibfnamefont {N.}~\bibnamefont {Vast}}, \bibinfo {author} {\bibfnamefont
  {X.}~\bibnamefont {Wu}}, \ and\ \bibinfo {author} {\bibfnamefont
  {S.}~\bibnamefont {Baroni}},\ }\bibfield  {title} {\enquote {\bibinfo {title}
  {Advanced capabilities for materials modelling with {Quantum} {ESPRESSO}},}\
  }\href {\doibase 10.1088/1361-648X/aa8f79} {\bibfield  {journal} {\bibinfo
  {journal} {Journal of Physics: Condensed Matter}\ }\textbf {\bibinfo {volume}
  {29}},\ \bibinfo {pages} {465901} (\bibinfo {year} {2017})}\BibitemShut
  {NoStop}%
\bibitem [{\citenamefont {Giannozzi}\ \emph {et~al.}(2020)\citenamefont
  {Giannozzi}, \citenamefont {Baseggio}, \citenamefont {Bonfà}, \citenamefont
  {Brunato}, \citenamefont {Car}, \citenamefont {Carnimeo}, \citenamefont
  {Cavazzoni}, \citenamefont {de~Gironcoli}, \citenamefont {Delugas},
  \citenamefont {Ferrari~Ruffino}, \citenamefont {Ferretti}, \citenamefont
  {Marzari}, \citenamefont {Timrov}, \citenamefont {Urru},\ and\ \citenamefont
  {Baroni}}]{giannozzi_2020}%
  \BibitemOpen
  \bibfield  {author} {\bibinfo {author} {\bibfnamefont {P.}~\bibnamefont
  {Giannozzi}}, \bibinfo {author} {\bibfnamefont {O.}~\bibnamefont {Baseggio}},
  \bibinfo {author} {\bibfnamefont {P.}~\bibnamefont {Bonfà}}, \bibinfo
  {author} {\bibfnamefont {D.}~\bibnamefont {Brunato}}, \bibinfo {author}
  {\bibfnamefont {R.}~\bibnamefont {Car}}, \bibinfo {author} {\bibfnamefont
  {I.}~\bibnamefont {Carnimeo}}, \bibinfo {author} {\bibfnamefont
  {C.}~\bibnamefont {Cavazzoni}}, \bibinfo {author} {\bibfnamefont
  {S.}~\bibnamefont {de~Gironcoli}}, \bibinfo {author} {\bibfnamefont
  {P.}~\bibnamefont {Delugas}}, \bibinfo {author} {\bibfnamefont
  {F.}~\bibnamefont {Ferrari~Ruffino}}, \bibinfo {author} {\bibfnamefont
  {A.}~\bibnamefont {Ferretti}}, \bibinfo {author} {\bibfnamefont
  {N.}~\bibnamefont {Marzari}}, \bibinfo {author} {\bibfnamefont
  {I.}~\bibnamefont {Timrov}}, \bibinfo {author} {\bibfnamefont
  {A.}~\bibnamefont {Urru}}, \ and\ \bibinfo {author} {\bibfnamefont
  {S.}~\bibnamefont {Baroni}},\ }\bibfield  {title} {\enquote {\bibinfo {title}
  {Quantum {ESPRESSO} toward the exascale},}\ }\href {\doibase
  10.1063/5.0005082} {\bibfield  {journal} {\bibinfo  {journal} {The Journal of
  Chemical Physics}\ }\textbf {\bibinfo {volume} {152}},\ \bibinfo {pages}
  {154105} (\bibinfo {year} {2020})}\BibitemShut {NoStop}%
\bibitem [{\citenamefont {Monkhorst}\ and\ \citenamefont
  {Pack}(1976)}]{monkhors_1976}%
  \BibitemOpen
  \bibfield  {author} {\bibinfo {author} {\bibfnamefont {H.~J.}\ \bibnamefont
  {Monkhorst}}\ and\ \bibinfo {author} {\bibfnamefont {J.~D.}\ \bibnamefont
  {Pack}},\ }\bibfield  {title} {\enquote {\bibinfo {title} {Special points for
  {Brillouin}-zone integrations},}\ }\href {\doibase 10.1103/PhysRevB.13.5188}
  {\bibfield  {journal} {\bibinfo  {journal} {Physical Review B}\ }\textbf
  {\bibinfo {volume} {13}},\ \bibinfo {pages} {5188--5192} (\bibinfo {year}
  {1976})}\BibitemShut {NoStop}%
\bibitem [{\citenamefont {von Barth}\ and\ \citenamefont
  {Gelatt}(1980)}]{von_barth_1980}%
  \BibitemOpen
  \bibfield  {author} {\bibinfo {author} {\bibfnamefont {U.}~\bibnamefont {von
  Barth}}\ and\ \bibinfo {author} {\bibfnamefont {C.~D.}\ \bibnamefont
  {Gelatt}},\ }\bibfield  {title} {\enquote {\bibinfo {title} {Validity of the
  frozen-core approximation and pseudopotential theory for cohesive energy
  calculations},}\ }\href {\doibase 10.1103/PhysRevB.21.2222} {\bibfield
  {journal} {\bibinfo  {journal} {Physical Review B}\ }\textbf {\bibinfo
  {volume} {21}},\ \bibinfo {pages} {2222--2228} (\bibinfo {year}
  {1980})}\BibitemShut {NoStop}%
\bibitem [{\citenamefont {Dal~Corso}\ \emph {et~al.}(1996)\citenamefont
  {Dal~Corso}, \citenamefont {Pasquarello}, \citenamefont {Baldereschi},\ and\
  \citenamefont {Car}}]{dal_corso_1996}%
  \BibitemOpen
  \bibfield  {author} {\bibinfo {author} {\bibfnamefont {A.}~\bibnamefont
  {Dal~Corso}}, \bibinfo {author} {\bibfnamefont {A.}~\bibnamefont
  {Pasquarello}}, \bibinfo {author} {\bibfnamefont {A.}~\bibnamefont
  {Baldereschi}}, \ and\ \bibinfo {author} {\bibfnamefont {R.}~\bibnamefont
  {Car}},\ }\bibfield  {title} {\enquote {\bibinfo {title}
  {Generalized-gradient approximations to density-functional theory: {A}
  comparative study for atoms and solids},}\ }\href {\doibase
  10.1103/PhysRevB.53.1180} {\bibfield  {journal} {\bibinfo  {journal}
  {Physical Review B}\ }\textbf {\bibinfo {volume} {53}},\ \bibinfo {pages}
  {1180--1185} (\bibinfo {year} {1996})}\BibitemShut {NoStop}%
\bibitem [{\citenamefont {Schlipf}\ and\ \citenamefont
  {Gygi}(2015)}]{schlipf_2015}%
  \BibitemOpen
  \bibfield  {author} {\bibinfo {author} {\bibfnamefont {M.}~\bibnamefont
  {Schlipf}}\ and\ \bibinfo {author} {\bibfnamefont {F.}~\bibnamefont {Gygi}},\
  }\bibfield  {title} {\enquote {\bibinfo {title} {Optimization algorithm for
  the generation of {ONCV} pseudopotentials},}\ }\href {\doibase
  10.1016/j.cpc.2015.05.011} {\bibfield  {journal} {\bibinfo  {journal}
  {Computer Physics Communications}\ }\textbf {\bibinfo {volume} {196}},\
  \bibinfo {pages} {36--44} (\bibinfo {year} {2015})}\BibitemShut {NoStop}%
\bibitem [{\citenamefont {Perdew}, \citenamefont {Burke},\ and\ \citenamefont
  {Ernzerhof}(1996)}]{perdew_1996}%
  \BibitemOpen
  \bibfield  {author} {\bibinfo {author} {\bibfnamefont {J.~P.}\ \bibnamefont
  {Perdew}}, \bibinfo {author} {\bibfnamefont {K.}~\bibnamefont {Burke}}, \
  and\ \bibinfo {author} {\bibfnamefont {M.}~\bibnamefont {Ernzerhof}},\
  }\bibfield  {title} {\enquote {\bibinfo {title} {Generalized {Gradient}
  {Approximation} {Made} {Simple}},}\ }\href {\doibase
  10.1103/PhysRevLett.77.3865} {\bibfield  {journal} {\bibinfo  {journal}
  {Physical Review Letters}\ }\textbf {\bibinfo {volume} {77}},\ \bibinfo
  {pages} {3865--3868} (\bibinfo {year} {1996})}\BibitemShut {NoStop}%
\bibitem [{\citenamefont {Ninomiya}\ and\ \citenamefont
  {Adachi}(1995)}]{ninomiya_1995}%
  \BibitemOpen
  \bibfield  {author} {\bibinfo {author} {\bibfnamefont {S.}~\bibnamefont
  {Ninomiya}}\ and\ \bibinfo {author} {\bibfnamefont {S.}~\bibnamefont
  {Adachi}},\ }\bibfield  {title} {\enquote {\bibinfo {title} {Optical
  properties of cubic and hexagonal {CdSe}},}\ }\href {\doibase
  10.1063/1.359815} {\bibfield  {journal} {\bibinfo  {journal} {Journal of
  Applied Physics}\ }\textbf {\bibinfo {volume} {78}},\ \bibinfo {pages}
  {4681--4689} (\bibinfo {year} {1995})}\BibitemShut {NoStop}%
\bibitem [{\citenamefont {Tancogne-Dejean}\ \emph {et~al.}(2020)\citenamefont
  {Tancogne-Dejean}, \citenamefont {Oliveira}, \citenamefont {Andrade},
  \citenamefont {Appel}, \citenamefont {Borca}, \citenamefont {Le~Breton},
  \citenamefont {Buchholz}, \citenamefont {Castro}, \citenamefont {Corni},
  \citenamefont {Correa}, \citenamefont {De~Giovannini}, \citenamefont
  {Delgado}, \citenamefont {Eich}, \citenamefont {Flick}, \citenamefont {Gil},
  \citenamefont {Gomez}, \citenamefont {Helbig}, \citenamefont {Hübener},
  \citenamefont {Jestädt}, \citenamefont {Jornet-Somoza}, \citenamefont
  {Larsen}, \citenamefont {Lebedeva}, \citenamefont {Lüders}, \citenamefont
  {Marques}, \citenamefont {Ohlmann}, \citenamefont {Pipolo}, \citenamefont
  {Rampp}, \citenamefont {Rozzi}, \citenamefont {Strubbe}, \citenamefont
  {Sato}, \citenamefont {Schäfer}, \citenamefont {Theophilou}, \citenamefont
  {Welden},\ and\ \citenamefont {Rubio}}]{tancogne-dejean_2020}%
  \BibitemOpen
  \bibfield  {author} {\bibinfo {author} {\bibfnamefont {N.}~\bibnamefont
  {Tancogne-Dejean}}, \bibinfo {author} {\bibfnamefont {M.~J.~T.}\ \bibnamefont
  {Oliveira}}, \bibinfo {author} {\bibfnamefont {X.}~\bibnamefont {Andrade}},
  \bibinfo {author} {\bibfnamefont {H.}~\bibnamefont {Appel}}, \bibinfo
  {author} {\bibfnamefont {C.~H.}\ \bibnamefont {Borca}}, \bibinfo {author}
  {\bibfnamefont {G.}~\bibnamefont {Le~Breton}}, \bibinfo {author}
  {\bibfnamefont {F.}~\bibnamefont {Buchholz}}, \bibinfo {author}
  {\bibfnamefont {A.}~\bibnamefont {Castro}}, \bibinfo {author} {\bibfnamefont
  {S.}~\bibnamefont {Corni}}, \bibinfo {author} {\bibfnamefont {A.~A.}\
  \bibnamefont {Correa}}, \bibinfo {author} {\bibfnamefont {U.}~\bibnamefont
  {De~Giovannini}}, \bibinfo {author} {\bibfnamefont {A.}~\bibnamefont
  {Delgado}}, \bibinfo {author} {\bibfnamefont {F.~G.}\ \bibnamefont {Eich}},
  \bibinfo {author} {\bibfnamefont {J.}~\bibnamefont {Flick}}, \bibinfo
  {author} {\bibfnamefont {G.}~\bibnamefont {Gil}}, \bibinfo {author}
  {\bibfnamefont {A.}~\bibnamefont {Gomez}}, \bibinfo {author} {\bibfnamefont
  {N.}~\bibnamefont {Helbig}}, \bibinfo {author} {\bibfnamefont
  {H.}~\bibnamefont {Hübener}}, \bibinfo {author} {\bibfnamefont
  {R.}~\bibnamefont {Jestädt}}, \bibinfo {author} {\bibfnamefont
  {J.}~\bibnamefont {Jornet-Somoza}}, \bibinfo {author} {\bibfnamefont {A.~H.}\
  \bibnamefont {Larsen}}, \bibinfo {author} {\bibfnamefont {I.~V.}\
  \bibnamefont {Lebedeva}}, \bibinfo {author} {\bibfnamefont {M.}~\bibnamefont
  {Lüders}}, \bibinfo {author} {\bibfnamefont {M.~A.~L.}\ \bibnamefont
  {Marques}}, \bibinfo {author} {\bibfnamefont {S.~T.}\ \bibnamefont
  {Ohlmann}}, \bibinfo {author} {\bibfnamefont {S.}~\bibnamefont {Pipolo}},
  \bibinfo {author} {\bibfnamefont {M.}~\bibnamefont {Rampp}}, \bibinfo
  {author} {\bibfnamefont {C.~A.}\ \bibnamefont {Rozzi}}, \bibinfo {author}
  {\bibfnamefont {D.~A.}\ \bibnamefont {Strubbe}}, \bibinfo {author}
  {\bibfnamefont {S.~A.}\ \bibnamefont {Sato}}, \bibinfo {author}
  {\bibfnamefont {C.}~\bibnamefont {Schäfer}}, \bibinfo {author}
  {\bibfnamefont {I.}~\bibnamefont {Theophilou}}, \bibinfo {author}
  {\bibfnamefont {A.}~\bibnamefont {Welden}}, \ and\ \bibinfo {author}
  {\bibfnamefont {A.}~\bibnamefont {Rubio}},\ }\bibfield  {title} {\enquote
  {\bibinfo {title} {Octopus, a computational framework for exploring
  light-driven phenomena and quantum dynamics in extended and finite
  systems},}\ }\href {\doibase 10.1063/1.5142502} {\bibfield  {journal}
  {\bibinfo  {journal} {The Journal of Chemical Physics}\ }\textbf {\bibinfo
  {volume} {152}},\ \bibinfo {pages} {124119} (\bibinfo {year}
  {2020})}\BibitemShut {NoStop}%
\bibitem [{\citenamefont {Vampa}\ \emph {et~al.}(2014)\citenamefont {Vampa},
  \citenamefont {McDonald}, \citenamefont {Orlando}, \citenamefont {Klug},
  \citenamefont {Corkum},\ and\ \citenamefont {Brabec}}]{vampa_2014}%
  \BibitemOpen
  \bibfield  {author} {\bibinfo {author} {\bibfnamefont {G.}~\bibnamefont
  {Vampa}}, \bibinfo {author} {\bibfnamefont {C.}~\bibnamefont {McDonald}},
  \bibinfo {author} {\bibfnamefont {G.}~\bibnamefont {Orlando}}, \bibinfo
  {author} {\bibfnamefont {D.}~\bibnamefont {Klug}}, \bibinfo {author}
  {\bibfnamefont {P.}~\bibnamefont {Corkum}}, \ and\ \bibinfo {author}
  {\bibfnamefont {T.}~\bibnamefont {Brabec}},\ }\bibfield  {title} {\enquote
  {\bibinfo {title} {Theoretical {Analysis} of {High}-{Harmonic} {Generation}
  in {Solids}},}\ }\href {\doibase 10.1103/PhysRevLett.113.073901} {\bibfield
  {journal} {\bibinfo  {journal} {Physical Review Letters}\ }\textbf {\bibinfo
  {volume} {113}},\ \bibinfo {pages} {073901} (\bibinfo {year}
  {2014})}\BibitemShut {NoStop}%
\bibitem [{\citenamefont {Müller}, \citenamefont {Sharma},\ and\ \citenamefont
  {Sierka}(2020)}]{muller_2020}%
  \BibitemOpen
  \bibfield  {author} {\bibinfo {author} {\bibfnamefont {C.}~\bibnamefont
  {Müller}}, \bibinfo {author} {\bibfnamefont {M.}~\bibnamefont {Sharma}}, \
  and\ \bibinfo {author} {\bibfnamefont {M.}~\bibnamefont {Sierka}},\
  }\bibfield  {title} {\enquote {\bibinfo {title} {Real‐time time‐dependent
  density functional theory using density fitting and the continuous fast
  multipole method},}\ }\href {\doibase 10.1002/jcc.26412} {\bibfield
  {journal} {\bibinfo  {journal} {Journal of Computational Chemistry}\ }\textbf
  {\bibinfo {volume} {41}},\ \bibinfo {pages} {2573--2582} (\bibinfo {year}
  {2020})}\BibitemShut {NoStop}%
\bibitem [{\citenamefont {Franzke}\ \emph {et~al.}(2023)\citenamefont
  {Franzke}, \citenamefont {Holzer}, \citenamefont {Andersen}, \citenamefont
  {Begušić}, \citenamefont {Bruder}, \citenamefont {Coriani}, \citenamefont
  {Della~Sala}, \citenamefont {Fabiano}, \citenamefont {Fedotov}, \citenamefont
  {Fürst}, \citenamefont {Gillhuber}, \citenamefont {Grotjahn}, \citenamefont
  {Kaupp}, \citenamefont {Kehry}, \citenamefont {Krstić}, \citenamefont
  {Mack}, \citenamefont {Majumdar}, \citenamefont {Nguyen}, \citenamefont
  {Parker}, \citenamefont {Pauly}, \citenamefont {Pausch}, \citenamefont
  {Perlt}, \citenamefont {Phun}, \citenamefont {Rajabi}, \citenamefont
  {Rappoport}, \citenamefont {Samal}, \citenamefont {Schrader}, \citenamefont
  {Sharma}, \citenamefont {Tapavicza}, \citenamefont {Treß}, \citenamefont
  {Voora}, \citenamefont {Wodyński}, \citenamefont {Yu}, \citenamefont
  {Zerulla}, \citenamefont {Furche}, \citenamefont {Hättig}, \citenamefont
  {Sierka}, \citenamefont {Tew},\ and\ \citenamefont {Weigend}}]{franzke_2023}%
  \BibitemOpen
  \bibfield  {author} {\bibinfo {author} {\bibfnamefont {Y.~J.}\ \bibnamefont
  {Franzke}}, \bibinfo {author} {\bibfnamefont {C.}~\bibnamefont {Holzer}},
  \bibinfo {author} {\bibfnamefont {J.~H.}\ \bibnamefont {Andersen}}, \bibinfo
  {author} {\bibfnamefont {T.}~\bibnamefont {Begušić}}, \bibinfo {author}
  {\bibfnamefont {F.}~\bibnamefont {Bruder}}, \bibinfo {author} {\bibfnamefont
  {S.}~\bibnamefont {Coriani}}, \bibinfo {author} {\bibfnamefont
  {F.}~\bibnamefont {Della~Sala}}, \bibinfo {author} {\bibfnamefont
  {E.}~\bibnamefont {Fabiano}}, \bibinfo {author} {\bibfnamefont {D.~A.}\
  \bibnamefont {Fedotov}}, \bibinfo {author} {\bibfnamefont {S.}~\bibnamefont
  {Fürst}}, \bibinfo {author} {\bibfnamefont {S.}~\bibnamefont {Gillhuber}},
  \bibinfo {author} {\bibfnamefont {R.}~\bibnamefont {Grotjahn}}, \bibinfo
  {author} {\bibfnamefont {M.}~\bibnamefont {Kaupp}}, \bibinfo {author}
  {\bibfnamefont {M.}~\bibnamefont {Kehry}}, \bibinfo {author} {\bibfnamefont
  {M.}~\bibnamefont {Krstić}}, \bibinfo {author} {\bibfnamefont
  {F.}~\bibnamefont {Mack}}, \bibinfo {author} {\bibfnamefont {S.}~\bibnamefont
  {Majumdar}}, \bibinfo {author} {\bibfnamefont {B.~D.}\ \bibnamefont
  {Nguyen}}, \bibinfo {author} {\bibfnamefont {S.~M.}\ \bibnamefont {Parker}},
  \bibinfo {author} {\bibfnamefont {F.}~\bibnamefont {Pauly}}, \bibinfo
  {author} {\bibfnamefont {A.}~\bibnamefont {Pausch}}, \bibinfo {author}
  {\bibfnamefont {E.}~\bibnamefont {Perlt}}, \bibinfo {author} {\bibfnamefont
  {G.~S.}\ \bibnamefont {Phun}}, \bibinfo {author} {\bibfnamefont
  {A.}~\bibnamefont {Rajabi}}, \bibinfo {author} {\bibfnamefont
  {D.}~\bibnamefont {Rappoport}}, \bibinfo {author} {\bibfnamefont
  {B.}~\bibnamefont {Samal}}, \bibinfo {author} {\bibfnamefont
  {T.}~\bibnamefont {Schrader}}, \bibinfo {author} {\bibfnamefont
  {M.}~\bibnamefont {Sharma}}, \bibinfo {author} {\bibfnamefont
  {E.}~\bibnamefont {Tapavicza}}, \bibinfo {author} {\bibfnamefont {R.~S.}\
  \bibnamefont {Treß}}, \bibinfo {author} {\bibfnamefont {V.}~\bibnamefont
  {Voora}}, \bibinfo {author} {\bibfnamefont {A.}~\bibnamefont {Wodyński}},
  \bibinfo {author} {\bibfnamefont {J.~M.}\ \bibnamefont {Yu}}, \bibinfo
  {author} {\bibfnamefont {B.}~\bibnamefont {Zerulla}}, \bibinfo {author}
  {\bibfnamefont {F.}~\bibnamefont {Furche}}, \bibinfo {author} {\bibfnamefont
  {C.}~\bibnamefont {Hättig}}, \bibinfo {author} {\bibfnamefont
  {M.}~\bibnamefont {Sierka}}, \bibinfo {author} {\bibfnamefont {D.~P.}\
  \bibnamefont {Tew}}, \ and\ \bibinfo {author} {\bibfnamefont
  {F.}~\bibnamefont {Weigend}},\ }\bibfield  {title} {\enquote {\bibinfo
  {title} {{TURBOMOLE}: {Today} and {Tomorrow}},}\ }\href {\doibase
  10.1021/acs.jctc.3c00347} {\bibfield  {journal} {\bibinfo  {journal} {Journal
  of Chemical Theory and Computation}\ }\textbf {\bibinfo {volume} {19}},\
  \bibinfo {pages} {6859--6890} (\bibinfo {year} {2023})}\BibitemShut {NoStop}%
\bibitem [{\citenamefont {Bannwarth}\ \emph {et~al.}(2021)\citenamefont
  {Bannwarth}, \citenamefont {Caldeweyher}, \citenamefont {Ehlert},
  \citenamefont {Hansen}, \citenamefont {Pracht}, \citenamefont {Seibert},
  \citenamefont {Spicher},\ and\ \citenamefont {Grimme}}]{bannwarth_2021}%
  \BibitemOpen
  \bibfield  {author} {\bibinfo {author} {\bibfnamefont {C.}~\bibnamefont
  {Bannwarth}}, \bibinfo {author} {\bibfnamefont {E.}~\bibnamefont
  {Caldeweyher}}, \bibinfo {author} {\bibfnamefont {S.}~\bibnamefont {Ehlert}},
  \bibinfo {author} {\bibfnamefont {A.}~\bibnamefont {Hansen}}, \bibinfo
  {author} {\bibfnamefont {P.}~\bibnamefont {Pracht}}, \bibinfo {author}
  {\bibfnamefont {J.}~\bibnamefont {Seibert}}, \bibinfo {author} {\bibfnamefont
  {S.}~\bibnamefont {Spicher}}, \ and\ \bibinfo {author} {\bibfnamefont
  {S.}~\bibnamefont {Grimme}},\ }\bibfield  {title} {\enquote {\bibinfo {title}
  {Extended tight‐binding quantum chemistry methods},}\ }\href {\doibase
  10.1002/wcms.1493} {\bibfield  {journal} {\bibinfo  {journal} {WIREs
  Computational Molecular Science}\ }\textbf {\bibinfo {volume} {11}},\
  \bibinfo {pages} {e1493} (\bibinfo {year} {2021})}\BibitemShut {NoStop}%
\bibitem [{\citenamefont {Grimme}, \citenamefont {Bannwarth},\ and\
  \citenamefont {Shushkov}(2017)}]{grimme_2017}%
  \BibitemOpen
  \bibfield  {author} {\bibinfo {author} {\bibfnamefont {S.}~\bibnamefont
  {Grimme}}, \bibinfo {author} {\bibfnamefont {C.}~\bibnamefont {Bannwarth}}, \
  and\ \bibinfo {author} {\bibfnamefont {P.}~\bibnamefont {Shushkov}},\
  }\bibfield  {title} {\enquote {\bibinfo {title} {A {Robust} and {Accurate}
  {Tight}-{Binding} {Quantum} {Chemical} {Method} for {Structures},
  {Vibrational} {Frequencies}, and {Noncovalent} {Interactions} of {Large}
  {Molecular} {Systems} {Parametrized} for {All} spd-{Block} {Elements} (
  \textit{{Z}} = 1–86)},}\ }\href {\doibase 10.1021/acs.jctc.7b00118}
  {\bibfield  {journal} {\bibinfo  {journal} {Journal of Chemical Theory and
  Computation}\ }\textbf {\bibinfo {volume} {13}},\ \bibinfo {pages}
  {1989--2009} (\bibinfo {year} {2017})}\BibitemShut {NoStop}%
\bibitem [{\citenamefont {Weigend}\ and\ \citenamefont
  {Ahlrichs}(2005)}]{weigend_2005}%
  \BibitemOpen
  \bibfield  {author} {\bibinfo {author} {\bibfnamefont {F.}~\bibnamefont
  {Weigend}}\ and\ \bibinfo {author} {\bibfnamefont {R.}~\bibnamefont
  {Ahlrichs}},\ }\bibfield  {title} {\enquote {\bibinfo {title} {Balanced basis
  sets of split valence, triple zeta valence and quadruple zeta valence quality
  for {H} to {Rn}: {Design} and assessment of accuracy},}\ }\href {\doibase
  10.1039/b508541a} {\bibfield  {journal} {\bibinfo  {journal} {Physical
  Chemistry Chemical Physics}\ }\textbf {\bibinfo {volume} {7}},\ \bibinfo
  {pages} {3297} (\bibinfo {year} {2005})}\BibitemShut {NoStop}%
\bibitem [{\citenamefont {Weigend}(2006)}]{weigend_2006}%
  \BibitemOpen
  \bibfield  {author} {\bibinfo {author} {\bibfnamefont {F.}~\bibnamefont
  {Weigend}},\ }\bibfield  {title} {\enquote {\bibinfo {title} {Accurate
  {Coulomb}-fitting basis sets for {H} to {Rn}},}\ }\href {\doibase
  10.1039/b515623h} {\bibfield  {journal} {\bibinfo  {journal} {Physical
  Chemistry Chemical Physics}\ }\textbf {\bibinfo {volume} {8}},\ \bibinfo
  {pages} {1057} (\bibinfo {year} {2006})}\BibitemShut {NoStop}%
\bibitem [{\citenamefont {Ikemachi}\ \emph {et~al.}(2017)\citenamefont
  {Ikemachi}, \citenamefont {Shinohara}, \citenamefont {Sato}, \citenamefont
  {Yumoto}, \citenamefont {Kuwata-Gonokami},\ and\ \citenamefont
  {Ishikawa}}]{ikemachi_trajectory_2017}%
  \BibitemOpen
  \bibfield  {author} {\bibinfo {author} {\bibfnamefont {T.}~\bibnamefont
  {Ikemachi}}, \bibinfo {author} {\bibfnamefont {Y.}~\bibnamefont {Shinohara}},
  \bibinfo {author} {\bibfnamefont {T.}~\bibnamefont {Sato}}, \bibinfo {author}
  {\bibfnamefont {J.}~\bibnamefont {Yumoto}}, \bibinfo {author} {\bibfnamefont
  {M.}~\bibnamefont {Kuwata-Gonokami}}, \ and\ \bibinfo {author} {\bibfnamefont
  {K.~L.}\ \bibnamefont {Ishikawa}},\ }\bibfield  {title} {\enquote {\bibinfo
  {title} {Trajectory analysis of high-order-harmonic generation from periodic
  crystals},}\ }\href {\doibase 10.1103/physreva.95.043416} {\bibfield
  {journal} {\bibinfo  {journal} {Physical Review A}\ }\textbf {\bibinfo
  {volume} {95}} (\bibinfo {year} {2017}),\ 10.1103/physreva.95.043416},\
  \bibinfo {note} {publisher: American Physical Society (APS)}\BibitemShut
  {NoStop}%
\end{thebibliography}%

\appendix

\section{Equivalence of the real-space periodic model to the semiconductor-Bloch-equations in  the moving-frame Wannier basis} \label{sec:sbe_equiv}
We define the density matrix in momentum space as
\begin{equation}
    \rho^{\mathbf{A}}_\mathbf{k} \equiv \sum\limits_{\mathbf{s}} \rho^{\mathbf{A}}_{\mathbf{s}} \mathrm{e}^{\mathrm{i}\mathbf{ks}}
                                   =  \sum\limits_{\mathbf{s}} \rho^{\mathbf{0}}_{\mathbf{s}} \mathrm{e}^{\mathrm{i} (\mathbf{k}+\mathbf{A})\mathbf{s}}
,
\end{equation}
where we used $\rho^{\mathbf{A}}_{\mathbf{s}} = \mathrm{e}^{\mathrm{i}\mathbf{As}} \rho^{\mathbf{0}}_{\mathbf{s}}$, which follows from equations (\ref{eq:defPhasedWannier}) and (\ref{eq:defPhasedRho}).
We use the following definitions of the Hamiltonian and dipole matrix elements in $\mathbf{k}$-space \cite{silva_2019}
\begin{equation} \begin{aligned}
    H^{\mathrm{W}, \mathbf{A}}_{\mathbf{k}} \equiv \sum\limits_{\mathbf{R}} \mathrm{e}^{\mathrm{i}(\mathbf{k}+\mathbf{A})\mathbf{R}} H_{\mathbf{R}} 
        = H^{\mathrm{W},\mathbf{0}}_{\mathbf{k} + \mathbf{A}}
    , \\
    \mathbf{D}^{\mathrm{W}, \mathbf{A}}_{\mathbf{k}} \equiv \sum\limits_{\mathbf{R}} \mathrm{e}^{\mathrm{i}(\mathbf{k}+\mathbf{A})\mathbf{R}} \mathbf{D}_{\mathbf{R}}
        = \mathbf{D}^{\mathrm{W},\mathbf{0}}_{\mathbf{k} + \mathbf{A}}
    ,
\end{aligned} \end{equation}
where we omitted the Wannier indices and applied the Fourier transform to eqs.~\eqref{eq:defWannierH} and \eqref{eq:defWannierR}, respectively.
We know define
\begin{equation}
    H^{\mathrm{W}, \mathbf{E}}_\mathbf{k} = H^{\mathrm{W},\mathbf{0}}_{\mathbf{k} + \mathbf{A}} + \mathbf{E}  \mathbf{D}^{\mathrm{W},\mathbf{0}}_{\mathbf{k} + \mathbf{A}}
,
\end{equation}
which is the Fourier transform of eq.~\eqref{eq:defHAE} in the moving frame basis.
Therefore, eq.~\eqref{eq:prop_periodic}, which is a convolution with respect to the lattice vectors, can be immediately transformed into
\begin{equation}
    \mathrm{i}\absDeriv{t} \rho_{\mathbf{k}+\mathbf{A}} = \left[ H^\mathrm{W}_{\mathbf{k}+\mathbf{A}}, \rho_{\mathbf{k}+\mathbf{A}}\right]
.
\end{equation}
These are the SBEs in the moving-frame Wannier basis \cite{silva_2019, yue_2022,thummler_semiconductor_2025}.

\end{document}